\documentclass[aps,pra,twocolumn,superscriptaddress,10pt,floatfix]{revtex4}
\usepackage[pdftex,plainpages=false,colorlinks=true,citecolor=blue,linkcolor=blue,urlcolor=blue,filecolor=green,bookmarksopen=true]{hyperref}
\usepackage{multirow}
\usepackage[utf8]{inputenc}
\usepackage[english]{babel}
\usepackage{amsmath}
\usepackage[caption = false]{subfig}
\usepackage{graphicx}
\usepackage{blindtext}
\usepackage{lipsum}
\usepackage{amsfonts}
\usepackage{bbm}
\usepackage{amssymb}
\usepackage{bbold}
\usepackage{enumerate}
\usepackage{float}
\usepackage{xcolor}
\usepackage{latexsym}

\usepackage{eucal}
\usepackage{times,txfonts}

\usepackage{bm}

\newcommand{\Ecal}{\mathcal{E}}

\newcommand{\1}{\mathbbm{1}}

\newcommand{\ket}[1]{\left| #1 \right\rangle}

\begin{document}

\title{Spin-orbit implementation of Solovay-Kitaev decomposition of single-qubit channels }

\author{M. H. M. Passos}
\email{mhmpassos@id.uff.br}
\affiliation{Instituto de Ciências Exatas, Universidade Federal Fluminense, Volta Redonda, Rio de Janeiro, Brazil}
\affiliation{Instituto de F\'{i}sica, Universidade Federal Fluminense, Niter\'{o}i, Rio de Janeiro, Brazil}

\author{A. de Oliveira Junior}
\email{alexssandre.oliveirajunior@uj.edu.pl}
\affiliation{Instituto de F\'\i sica  Gleb Wataghin,  Universidade Estadual de Campinas, Campinas, SP, Brazil} 
\affiliation{Faculty of Physics, Astronomy and Applied Computer Science, Jagiellonian University, 30-348 Kraków, Poland} 

\author{M. C. de Oliveira}
\email{marcos@ifi.unicamp.br}
\affiliation{Instituto de F\'\i sica  Gleb Wataghin,  Universidade Estadual de Campinas, Campinas, SP, Brazil} 

\author{A. Z. Khoury}
\email{azkhoury@id.uff.br}
\affiliation{Instituto de F\'{i}sica, Universidade Federal Fluminense, Niter\'{o}i, Rio de Janeiro, Brazil}

\author{J. A. O. Huguenin}
\email{jose\_huguenin@id.uff.br}
\affiliation{Instituto de Ciências Exatas, Universidade Federal Fluminense, Volta Redonda, Rio de Janeiro, Brazil}
\affiliation{Instituto de F\'{i}sica, Universidade Federal Fluminense, Niter\'{o}i, Rio de Janeiro, Brazil}

\begin{abstract}
The Solovay-Kitaev theorem allows us to approximate any single-qubit gate to arbitrary accuracy with a finite sequence of fundamental operations from a universal set of gates. Inspired by this decomposition,  we present a quantum channel simulator capable of implementing any completely positive trace-preserving map. Our realization consists of one ancillary qubit, encoded in the transverse mode of a laser beam (orbital degree of freedom), one qubit system, encoded in its polarization (spin), one spin-orbit CNOT gate and four single-qubit operations performed with prisms and polarization components. Our results describe the implementation of arbitrary single-qubit channels on the photon polarization using the transverse mode as the ancillary qubit.
\end{abstract}

\maketitle

\section{Introduction}

Quantum channels are completely positive (CP) trace-preserving maps between operator spaces, allowing transmission of both classical and quantum information. Any quantum channel is ultimately implemented on a physical system and therefore is subjected to external noise. In contrast to the simplicity of the binary channel in classical communication \cite{CoverThomas}, there are several ways in which the state of a quantum bit can be affected when communicated over a noisy channel \cite{Nielsen:Book}. It is considerably challenging to devise a simple experimental procedure allowing the implementation of the several effects that an arbitrary channel may impose on an encoding qubit. This is particularly relevant in quantum thermodynamics, where it is fundamental to the simulation of controllable reservoirs.  Historically, optical implementations of relevant quantum channels, as amplitude damping, phase-damping, bit flip channels, among others, were performed by using single photons, (e.g. ~\cite{PRA.78.Davidovich}). On the other hand, the degrees of freedom of intense laser beams have been widely employed to simulate single-photon experiments, and the results show that such a platform is extremely convenient as a test-bed for several quantum properties in a rather simple way~\cite{PRL.99.Topo,Eberly}. 
Indeed, it can be shown that such systems can be used to observe violations of quantum-like inequalities~\cite{PRA.82.Borges,NaturePhot.Kagalwala-2012,Opt.Lett.Balthazar-2016}. Moreover, many other quantum protocols can be investigated, such as quantum key distribution~\cite{PRA.77.Cadu-Cripto}, teleportation~\cite{PRA.83.Zela-Teleport} and quantum logical gates~\cite{Opt.Exp.Souza-Cnot-2010,JOPS-B.Cod.Op.Balthazar-2016}. As a further implementation of interest here, it is essential to highlight the experimental simulation of open quantum systems to investigate environment-induced entanglement~\cite{PRA.97.Enviro-Passos}. None of those constituted a systematic channel implementation procedure, though.

Alternatively, one could use the well-known fact that an arbitrary unitary operation $U$ can be implemented through a circuit consisting of single-qubit operations, auxiliary qubits and controlled-\textsl{NOT} (CNOT) gates. Such universality is essential since it guarantees the equivalence of possibly different models of quantum computation. For example, we may design a quantum circuit comprising of four input and output qubits and simulate it with a constant number of CNOT and single-qubit unitary gates. However, while the single-qubit gates form a continuum, the methods for fault-tolerant quantum computation \cite{Preskill} works only for a discrete set of gates. Fortunately, the celebrated \emph{Solovay-Kitaev} theorem \cite{Kitaev_1997} addresses this problem, stating that any unitary operation $U$ can be approximated using a fixed finite set of gates. Dawson and Nielsen \cite{DN} introduced an algorithm for the Solovay-Kitaev decomposition, and more recently, inspired by this decomposition, a method for approximating an arbitrary single-qubit channel using single-qubit gates and a controlled-NOT was proposed in Ref.\cite{PhysRevLett.111.130504}. Since then, many alternative methods for simulation of general qubit and qudit channels (see e.g. \cite{Wang_2015}) were proposed, and there were several attempts on the experimental implementation of quantum channels \cite{Hu-Exp_SK:18,PhysRevA.96.062303,McCutcheon-EXP_SK:18}. Particularly relevant for the present discussion is the proposal of Ref.\cite{PhysRevLett.111.130504}, implemented in Ref.\cite{SK-Lu_Wei:17} using photon pairs generated by spontaneous parametric down-conversion. It is certainly relevant to extend the implementation of the Solovay-Kiataev decomposition to other systems, where different conditions may apply. 

In that sense spin-orbit modes have been demonstrated to be a powerful platform with great potential for applications in quantum information science \cite{Marrucci12,Cardano12,Cardano01,Karimi12,Nagali09,DeOliveira20,Goyal13,Hamadou13,Konrad19,McLaren15,Gailele18,Johnson19,Mirhosseini15,Mirhosseini16,Arlt99}, especially in optical communication systems, for which was recently presented important technological proposals \cite{PRA.77.Cadu-Cripto, Thomaschewski,Gregg,Karimi,Cardano,marrucci:2012}. Therefore, the study of arbitrary quantum channels with the spin-orbit modes is extremely useful to improve the acknowledgment about quantum communication protocols that employ structured light.

The transverse mode structure of the electromagnetic field constitutes an infinite-dimensional Hilbert space that combines with the photon polarization in a tensor product space. This extra degree of freedom is associated with the photon orbital angular momentum (OAM) given by discrete values $L_{orb} = m\hbar$ ($m\in\mathbb{Z}$). When restricted to the first order subspace $(|m|=1)$, the transverse modes exhibit a qubit structure that can be combined with polarization to encode a two-qubit Hilbert space in every single photon. In this context, entanglement is manifested as polarization vortices that constitute non-separable spin-orbit modes. This two-qubit encoding on every single photon is particularly convenient for our purpose since it allows for an easy cross-talk between the two degrees of freedom. In this way, several local and controlled operations can be readily performed with the aid of adequate prisms and polarization optics. It constitutes a quite versatile platform for implementing arbitrary single-qubit channels on the photon polarization using the transverse mode as the ancillary qubit. Moreover, the channel characterization is based on probability measurements that are readily obtained from the output intensities in the experimental apparatus. Therefore, we could perform our measurements with inexpensive laser sources and detectors, without the need for single-photon modules.

In this paper, we present the construction of a quantum channel simulator capable of implementing any completely positive trace-preserving map. Our realization consists in one ancillary qubit encoded in the transverse mode of a laser beam (orbital degree of freedom), one system qubit encoded in its polarization (spin), one spin-orbit CNOT gate and four single-qubit operations performed with prisms and polarization components. The paper is organized as follows, in Section \ref{sec:background} we present the basic theoretical tools for describing arbitrary channels of single qubits, in Section \ref{sec:ExperimentalImplementation} we describe the experimental procedures for implementing arbitrary quantum channels, in Section \ref{sec:Results} we show the experimental results for several relevant quantum channels. Finally, our conclusions are drawn in Section \ref{sec:conclusions}.

\section{Background}\label{sec:background}

The remarkable Solovay-Kitaev theorem \cite{Nielsen:Book, Kitaev} provides a systematic procedure for approximating arbitrary unitary operations using a finite set of gates. In this context, the approach given by \cite{PhysRevLett.111.130504} has shown how to implement an arbitrary single-qubit channel using a CNOT and a universal set of single-qubit gates. This approach is appealing from the experimental point of view since only two qubits are required for its implementation. In what follows, we sketch the theoretical background, and the formalism is established.

We begin by considering a single qubit system $\rho \in \mathcal{T} (\mathcal{H}^{\textrm{S}})$ with $\mathcal{H}^{\textrm{S}}$ being a two-dimensional Hilbert space and $\mathcal{T}(\mathcal{H})$ denoting the set of operators on Hilbert space $\mathcal{H}$. An arbitrary  channel $\Ecal: \mathcal{D}(\mathcal{H}^{\textrm{S}}) \rightarrow \mathcal{D}(\mathcal{H}^{\textrm{S}})$ acting on a density operator,
\begin{equation}\label{eq:densOpt}
\rho = \frac{1}{2}(I + \mathbf{r} \cdot \pmb{\sigma} ), 
\end{equation}
where $\pmb{\sigma} = (\sigma_x, \sigma_y,\sigma_z)$ and $I$ is the $2\times 2$ identity matrix, must be completely positive trace-preserving (CPTP). A general way of describing a CP map is in terms of the Kraus operators $\{ K_i\}$ \cite{Book:Krauss}:
\begin{equation}\label{eq:1qbitMap}
 \Ecal (\rho) = \sum_i K_i \rho K^{\dagger}_i \, ,
\end{equation}
which form a linearly independent set, with the trace-preserving condition $\sum_i K^\dagger_i K_i = I$. The analysis for a general single-qubit CPTP map can be recast in terms of the geometric description, in which 
\begin{equation}
\Ecal(\rho) = \frac{1}{2}(I + \mathbf{r}' \, \cdot \, \pmb{\sigma}) \; , \quad \mathbf{r}{'} = T\textbf{r} + \mathbf{t} \, ,\label{eqre}
\end{equation}
encompassing a change on the Bloch vector $\mathbf{r}$, given by the distortion $T$-matrix, and a displacement, given by the vector $\mathbf{t}$.
 
The $T$-matrix  can be written into a diagonal form via a singular-value decomposition \cite{BETHRUSKAI2002159} for quasiextreme channels,
\begin{equation}
T=\text{diag}\left(\cos\nu,\cos\mu,\cos\nu\cos\mu\right),\label{tdiag}\end{equation}
in which case $\mathbf{t}=(0,0,\sin\nu\sin\mu)^T$ \cite{PhysRevLett.111.130504}.
This decomposition can be obtained in terms of a sum of positive operators representation, given only by two Kraus operators,
\begin{equation}\label{eq:KrausOp}
    K_0 =  \left(
\begin{array}{cc}
\cos\;\beta & 0\\
0 & \cos\;\alpha\\
\end{array}
\right), \;\;\;\;\;
  K_1 =  \left(
\begin{array}{cc}
0 & \sin\;\alpha\\
\sin\;\beta & 0\\
\end{array}
\right),
\end{equation}
with $\alpha = (\mu+\nu)/2$ and $\beta = (\mu-\nu)/2$.

\begin{figure}[!htb]
 \begin{center}
 \includegraphics[scale=0.65]{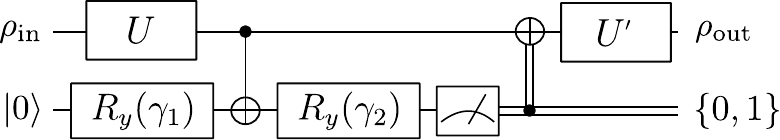} 
\end{center}
\caption{The circuit to implement an arbitrary channel $\Ecal$ (\ref{qechannel}) on a single qubit. The circuit implements each one of the channels $\Ecal^e_1$ and $\Ecal^e_2$, individually.}
 \label{fig:Circuit-SL}
\end{figure}

Remarkably, as it was proposed in Ref.\cite{PhysRevLett.111.130504}, any single-qubit CPTP channel $\Ecal$ can be decomposed into the convex combination 
\begin{equation}
\label{qechannel}
\Ecal = p \Ecal^e_a + (1-p)\Ecal^e_b,
\end{equation}
with $0 \leq p \leq 1$, and be simulated with only one ancillary qubit, two CNOTs and four single-qubit operations. Here $\Ecal^e_i,\,\, i=a,b$, are two realizations of quasiextreme channels (\ref{eqre}) represented in terms of the new Kraus operators,
\begin{equation}\label{eq:GEMap_Kraus}
    M_i = U\, K_i\, U', \;\;\; 
\end{equation}
where $\{U,U'\}\in$ SU(2), so that 
\begin{eqnarray}
U = \left(\begin{array}{cc}
u & -w^*\\
w & u^*\\
\end{array}
\right)\,,
\label{eq:Uuw}
\end{eqnarray}
with $|u|^2 + |w|^2 = 1\,$. Therefore, a general SU(2) operation can be represented by three independent real parameters on the hypersphere $S^3$, and there are different types of parametrization. For example, it can be characterized as a rotation by an angle $\psi\in [0,\pi]$ around a unit vector $\mathbf{\hat{n}}$ oriented along the direction given by the sagittal ($\theta\in [0,\pi]$) and azimuthal ($\phi\in [0,2\pi]$) angles, which gives
\begin{eqnarray}
U &=& R_{\mathbf{\hat{n}}}(\psi) = e^{-i\,\psi\,\mathbf{\hat{n}}\,\cdot\,\pmb{\sigma}}
=\cos\psi\,I - i\sin\psi\, \mathbf{\hat{n}}\cdot\pmb{\sigma}\;,
\nonumber\\
u &=& \cos\psi - i\cos\theta\sin\psi\;,
\nonumber\\
w &=& -i\sin\theta\sin\psi \,e^{i\phi}\;.
\label{eq:UR}
\end{eqnarray}
Another possible parametrization uses the Euler angles, which is suitable for optical implementations 
with polarization components.  In this parametrization, a general SU(2) matrix can be written as a 
sequence of three rotations characterized by the Euler angles $(\varphi,\xi,\zeta)\,$,
\begin{eqnarray}
&&    U(\varphi,\xi,\zeta) = R_y(\varphi)\, R_z(-\xi)\, R_y(\zeta)\;,
    \nonumber\\
&&    u = \cos\xi\cos(\varphi+\zeta) + i \sin\xi\cos(\varphi-\zeta)\;,
    \nonumber\\
&&    w = \cos\xi\sin(\varphi+\zeta) + i \sin\xi\sin(\varphi-\zeta)\;.
\label{SU2Euler}
\end{eqnarray}
It provides a simple relationship between the group parameters and the 
orientations of retardation devices such as half- and quarter-waveplates used for polarization transformations 
or mode converters used for transverse mode operations. We will give these relations explicitly in the experimental realization section.   
The circuit to be implemented is shown in Fig.\ref{fig:Circuit-SL}. It involves two SU(2) operators, $U$ and $U'$, which diagonalize the distortion matrix, two local operations $R_y(\gamma_{1,2})$ acting on the ancilla, two controlled operations and one measurement on the ancilla. The first CNOT gate is controlled by the system qubit and the second one acts on the system qubit conditioned to the ancilla measurement. 

In order to evaluate the implementation of the SK-decomposition in spin-orbit modes, the final state is characterized by reconstructing the density operator over a state tomography process, and the respective fidelity is obtained \cite{Hu-Exp_SK:18}. The effectiveness of our proposal can also be explored using state properties, such as how the quantum coherence \cite{Streltsov:18} is affected by a Markovian process during its propagation through the channel \cite{Passos_Markov}.

Quantum coherence is the central building block of quantum physics, yielding essential aspects of the principle of superposition in quantum computing, quantum teleportation, and many others that stem from this fundamental behaviour of the quantum domain. A frequently used definition to coherence is the $l_1$-norm quantum coherence \cite{Baumgratz:14,Streltsov:17}. By considering the general form of the density operator in (\ref{eq:densOpt})
we can 
write the $l_1-$norm coherence for a qubit state as \cite{Streltsov:18}
\begin{equation}\label{eq:Coh_l1}
C_{l_1}(r_x,r_y) = \sqrt{r_x^2 + r_y^2} = C_{l_1} = 2|\rho_{12}(t)|.
\end{equation}
The above definition is base-dependent and may present some changes with respect to local unitary operations. A basis-free quantum coherence measure was proposed in Ref.\cite{Streltsov:18}. The idea consists in maximizing the coherence over all local unitary transformations, resulting in
\begin{equation}\label{eq:CohMax}
C_{max}(r) = r,
\end{equation}
where $r=|\mathbf{r}|$ is the modulus of the Bloch vector. Performing tomographic measurements on the output state, both $l_1-$norm and maximal coherence are obtained and can be contrasted with the evolution of the decomposed channel state coherence.

\section{Experimental implementation} \label{sec:ExperimentalImplementation}

\begin{figure}[!htb]
 \begin{center}
 \includegraphics[scale=0.40,clip,trim=0mm 0mm 0mm 0mm]{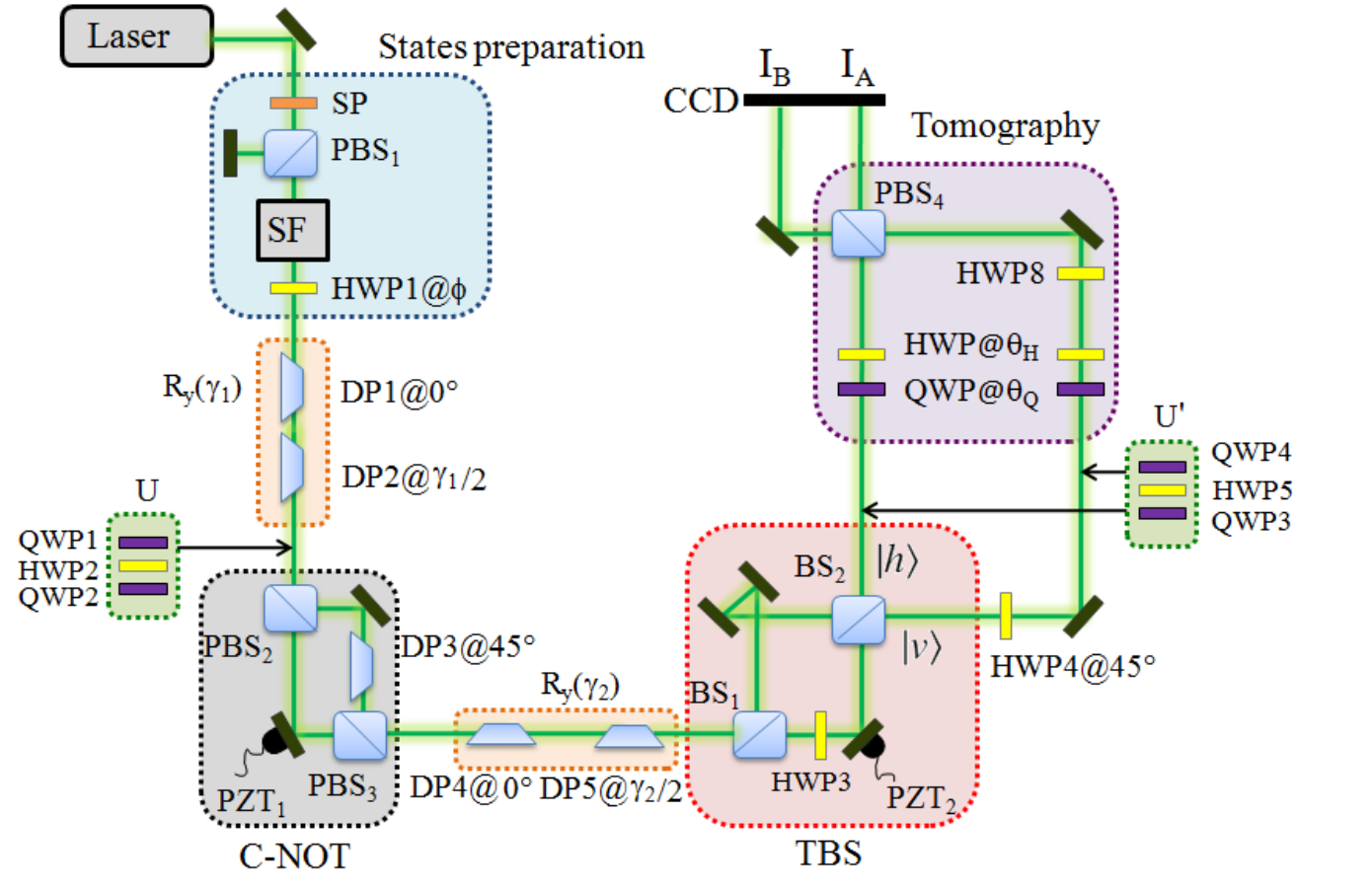} 
\end{center}
\caption{Experimental setup. SP stands for S-wave plate, SF for the spatial filter, HWP for the half-wave plate, QWP for the quarter-wave plate, BS for beam splitter, PBS for polarized beam splitter, and CCD for charge-coupled device camera.}
 \label{Fig:ExpSetup}
\end{figure}

The spin-orbit implementation of the Solovay-Kitaev decomposition was performed by the optical circuit illustrated in Fig.\ref{Fig:ExpSetup}. The system (upper wire in Fig.\ref{fig:Circuit-SL}) was encoded in the polarization degree of freedom of a laser beam ($@532~nm; 1.5~mW$) whereas the ancilla (lower wire in Fig.\ref{fig:Circuit-SL}) was encoded in the first-order transversal modes. For the qubit system, the horizontal polarization state $\ket{H}$ represents the ground state $\ket{0}$, and the vertical polarization $\ket{V}$  represents the excited state $\ket{1}$. Similarly, the transverse mode HG$_{10}$ (labeled  $\ket{h}$) denotes the ground state $\ket{0}$ for the ancilla, while the transverse mode HG$_{01}$ (labeled $\ket{v}$) represents the excited state $\ket{1}$.

Let us discuss the operations required for the SK-decomposition. For the qubit system, general transformations $U$ and  $U'$ are required, and for the polarization degree of freedom, the most general transformation is obtained by using two quarter-wave plates (QWP) and one half-wave plate (HWP), disposed in the following order QWP($\eta_2$), HWP($\tau$), QWP($\eta_1$), where the arguments are the respective angles of the fast axis with respect to the horizontal direction. According to the Jones matrix representation \cite{Jones-i:41}, the most general SU(2) operator can be decomposed as a product of waveplate operations in the following way \cite{Simon:90,Reddy:14}
\begin{eqnarray}\label{eq:QWP+HWP+QWP-Matrix}
   U &=& \text{QWP}(\eta_1) \, \text{HWP}(\tau)\, \text{QWP}(\eta_2)\;,
\end{eqnarray}
where 
\begin{eqnarray}\label{eq:QWP+HWP+QWP-Matrix2}
\text{QWP}(\eta) &=& R_y(\eta)\,Q_0\,R_y(-\eta)\;,
\nonumber\\
\text{HWP}(\tau) &=& R_y(\tau)\,H_0\,R_y(-\tau)\;,
\nonumber\\
Q_0 &=&\left(
\begin{array}{cc}
1 & \,\,\,0\\
0 & \,\,\,i\\
\end{array}
\right)\;,
\nonumber \\ 
H_0 &=& \left(
\begin{array}{cc}
1 & \,\,\,0\\
0 & -1\\
\end{array}
\right)\;.
\end{eqnarray}
Note that although $\text{QWP}(\eta)$ and $\text{HWP}(\tau)$ are not SU(2) matrices, the product 
given in \eqref{eq:QWP+HWP+QWP-Matrix} belongs to SU(2). Carrying out the matrix product, we can 
identify the complex numbers of the general form given in \eqref{eq:Uuw}
\begin{eqnarray}\label{eq:Coef.QHQ-Matrix}
&& u = \cos\Lambda \cos\eta_- - i \sin\Lambda \sin\eta_+ \,,
\nonumber \\
&& w = \cos\Lambda \sin\eta_- + i \sin\Lambda \cos\eta_+ \,,
\nonumber\\
&& \eta_{\pm} = \eta_1 \pm \eta_2 \,, 
\nonumber\\
&&\Lambda = 2\tau - \eta_1 - \eta_2\,.
\end{eqnarray}
Therefore, a simple relation can be established between the waveplates and the Euler angles
\begin{eqnarray}
    \eta_1 &=& \varphi - \pi/4\;,
\nonumber\\
    \eta_2 &=& -\zeta - \pi/4\;,
\nonumber\\
    \tau &=& (\varphi + \xi -\zeta)/2 - \pi/4\;.
\end{eqnarray}
Then, by choosing appropriately the angles $\tau$, $\eta_1$, and $\eta_2$, we can implement the general transformations, $U$ and $U'$, necessary to compose the operators $M_i$ of the SK-decomposition.

For the ancilla qubit we need to implement the operator $R_y (\gamma_{1,2})\,$. This transformation has to be performed in the transverse modes $\ket{h}$ and $\ket{v}$, and it is implemented by combining a sequence of two Dove prisms. The action of a Dove prism is the transverse mode analogous to a half-wave plate. Therefore, its Jones matrix is given by
\begin{equation}
\text{DP}(\gamma) = R_y(\gamma)\,H_0\,R_y(-\gamma) = 
\left(
\begin{array}{cc}
\cos 2\gamma & \sin 2\gamma\\
\sin2\gamma & -\cos 2\gamma\\
\end{array}
\right).
\end{equation}
Since $H_0\,R_y(-\gamma) = R_y(\gamma)\,H_0$ and $H_0^2 = I\,$, as can be easily verified, 
the ancilla rotations can be simply realized by a sequence of two Dove prisms, one 
horizontally oriented and the other rotated by $\gamma/2\,$, so that 
\begin{equation}\label{Eq:Exp-R(gamma)}
    R_y(\gamma) = \text{DP}(\gamma/2)\,\text{DP}(0)\;.
\end{equation}

With all the required transformations for the degrees of freedom representing the system and ancilla qubit, the optical circuit used to implement the Solovay-Kitaev decomposition in spin-orbit modes is sketched in Fig.\ref{Fig:ExpSetup}. A diode-pumped solid-state (DPSS) laser beam (532nm, 1.5mw power, horizontally polarized) goes to a S-wave plate (SP) to produce a state $(\ket{Hh}+\ket{Vv})/\sqrt{2}$. As we are interested in producing an initial state in which the system and the ancilla are in the ground state, the component $\ket{Hh}$ of the state produced by SP is selected. Thus, a polarized beam splitter (PBS$_1$) is introduced such that the component $\ket{Hh}$ will be transmitted, whereas the component $\ket{Vv}$ will be reflected and blocked. Then, the selected state $\ket{Hh}$ goes to a spatial filter (SF) to improve the mode fidelity. It is worth mentioning that we considered using an S-wave plate to produce the initial state due to its good fidelity in producing the desired mode. After the SF, the state $\ket{Hh}$ goes to a half-wave plate HWP1 with its fast axis rotated by an angle $\phi$ with respect to the horizontal axis, allowing us to produce any initial linear polarization state. Therefore, our initial state $\ket{\psi_i}$ is written as
\begin{equation}\label{eq:InitialSatate}
\ket{\psi_i} = \left(\cos 2\phi \ket{H} + \sin 2\phi \ket{V} \right) \otimes \ket{h}.
\end{equation}

It is important to note that we chose to work with pure states, whose evolution in the channels can be simply followed by the respective density matrices. Furthermore, as coherence is another parameter analyzed to verify the channel's implementation, and it is calculated from the absolute value of the non-diagonal elements of $\rho$, we did not care about a relative phase. Even though we do not have the most general initial state, the experimentally produced polarization state has the necessary ingredients to verify the main actions of the decomposed channels.

After $\ket{\psi_i}$ preparation, the next step is to implement the rotation $R_y(\gamma_1)$ in the ancilla qubit (see Fig.\ref{fig:Circuit-SL}). Hence, the laser beam is sent through a Dove prisms DP1 oriented at $0^{\circ}$ and DP2 oriented at $\gamma_1/2$. Depending on the channel we are interested in performing, we must implement the transformation $U'$ in the qubit system. Thus, the set QWP1, HWP2 and QWP2 oriented at angles $\eta_1\,$, $\tau$ and $\eta_2\,$, respectively, can be inserted in the laser path. Note that this transformation on polarization states will only be implemented in the bit phase flip channel.

The next stage showed in the circuit illustrated in Fig.\ref{fig:Circuit-SL} is the operation of a CNOT gate using a qubit as control. To implement this operation, we consider an interferometer composed by two polarized beam splitter PBS$_2$ and PBS$_3$, a Dove prism (DP3) at an angle of 45$^{\circ}$, and a mirror mounted on a piezoelectric ceramic (PZT), which is responsible for controlling the optical path difference in order to superpose coherently the states coming from the two arms of the interferometer. The polarization state is the control qubit, whereas the transverse mode is the target qubit. In this way, the state $\ket{H}$ is transmitted by the PBS$_2$, and the transverse mode does not suffer any transformation, leaving the CNOT  transmitted by the PBS$_3$. On the other hand, the state $\ket{V}$  is reflected by the PBS$_2$ and the Dove prism DP3 implements the following transformations $\ket{h} \rightarrow \ket{v}$ and $\ket{v} \rightarrow \ket{h}$ on the transverse mode. After these transformations, the state goes to the PBS$_3$ where it is reflected and coherently superposed to the state horizontally polarized that arrives through the other port of PBS$_3\,$. With these operations, the CNOT gate is completed. It is important to stress that in our CNOT implementation, target and control qubits are encoded in two independent degrees of freedom of a single photon, while in CNOT gates performed in photon pairs, for example, the qubits are spatially separated \cite{SK-Lu_Wei:17,Huang18,Huang17,Lu07}. No significant difference in the CNOT performance is expected once we can act independently on both degrees of freedom.

After the CNOT gate, the laser beam passes through DP4 oriented at $0^{\circ}$ and DP5 at $\gamma_2/2\,$, 
which implement another rotation in the ancilla given by the operator $R_y(\gamma_2)\,$.

The next step presented in Fig.\ref{fig:Circuit-SL} is the projective measurements in the ancilla. This step can be accomplished with the aid of a Mach-Zehnder interferometer with an additional mirror (MZIM) that sorts polarization and transverse modes of even and odd parities \cite{MZIM_Sassada}. The MZIM is sensitive to the combined parity of polarization and transverse modes. Its functionality is based on the extra $\pi$ phase acquired by modes $\ket{Hv}$ and $\ket{Vh}$ at the extra reflection arm. When a a half-wave plate oriented at $0^\circ$ is inserted in one arm of the MZIM, the device becomes polarization independent and can be used to sort transverse modes $\ket{h}$ and $\ket{v}$ regardless to the polarization state. Such an arrangement is considered a transverse beam splitter (TBS) - see the red dashed board in Fig.\ref{Fig:ExpSetup}. Two beam splitters compose the device, 50/50 BS$_1$ and BS$_2$, a half-wave plate (HWP3 at $0^{\circ}$) and a piezoelectric ceramic (PZT$_2$) to control the coherent superposition at the TBS output such that the horizontal component $\ket{h}$ of the transverse mode always leaves through port O$_1$ while the vertical component $\ket{v}$ leaves through O$_2$. More details are presented in Appendix \ref{Ap:MZIM+HWP}. 

Following the Solovay-Kitaev circuit in Fig.\ref{fig:Circuit-SL}, we need to implement a CNOT gate by choosing the ancilla as control and the qubit as the target. Our circuit is realized by performing a $\sigma_x$ operation on the polarization state at the $\ket{v}$ output of the TBS, implemented by the half-wave plate HWP4 oriented at $45^{\circ}$. After the $\sigma_x$ operation, as can be seen in Fig.\ref{fig:Circuit-SL}, we perform the last transformation $U'$ in the qubit system depending on the desired channel. In our setup, it is implemented by the set QWP3, HWP5 and QWP4 oriented at $\eta^\prime_2\,$, $\tau^\prime$ and $\eta^\prime_1\,$, respectively, placed at both TBS outputs. Then, the SK-decomposition is resumed.

In order to characterize the output state of the system $\rho_{out}$, a tomographic measurement \cite{Photonic-State-Tomography,Passos_Machine} of the state polarization is performed as sketched in the dashed box named tomography. The measurement in the three bases can be performed by the set composed by the QWP at $\theta_{Q}$, HWP at $\theta_{H}$, and PBS$_4$. By setting $\theta_{Q}=\theta_{H}=0$, the measurement basis is set to \{$\ket{H},\ket{V}$\}. With $\theta_{Q}= 45^\circ$ and $\theta_{H}=22.5$, we are able to measure in the diagonal basis \{$\ket{+},\ket{-}$\}, and for $\theta_{Q}= 45^\circ$ and $\theta_{H}=0$ the measurement is performed in the left-right basis \{$\ket{L},\ket{R}$\}. Note that HWP8 at $45^{\circ}$ is used to combine
the $\ket{H}$ polarization state leaving both outputs of the TBS at the output I$_A$ of PBS$_4$ and 
the $\ket{V}$ polarization state at the output I$_B\,$. The resulting intensities are projected on a screen and captured in a single image by a charge-coupled device (CCD) camera.

The matrix reconstruction is obtained through the Stokes parameters that relate the measurement statistics with the parameters of Eq. \eqref{eq:densOpt} in the following way \cite{Photonic-State-Tomography}
\begin{eqnarray}\label{eq:rhoS-exp}
r_x &=& P_{\ket{+}}-P_{\ket{-}}\;,
\nonumber\\
r_y &=& P_{\ket{L}}-P_{\ket{R}}\;,
\\
r_z &=& P_{\ket{H}}-P_{\ket{V}}\;.
\nonumber
\end{eqnarray}
The purity of the reconstructed state is given by $0\leq \|\mathbf{r}\| \leq 1\,$.
In our experiment, the normalized intensities recorded by the CCD camera  play the role of the probabilities. In this setup, I$_A$ is associated with the intensities of components $H,~ +~,L$  and I$_B$ related to the intensities of the components $V,~-~,R$. Thereby, the probability $P_{\ket{A}}$, of measuring a state $\ket{A} \equiv {\ket{H}, \ket{+}, \ket{L}}$ is given by
\begin{equation}\label{eq:Prob_Int}
P_{\ket{A}} = \frac{I_{\ket{A}}}{I_{\ket{A}}+I_{\ket{B}}}\;.
\end{equation}
The probability of measuring a state $\ket{B} \equiv {\ket{V}, \ket{-}, \ket{R}}$ is simply given by $P_{\ket{B}}=1-P_{\ket{A}}$\,. Next, we discuss the implementation of the decomposition for some important channels. For all implementations, the following values of the decoherence parameter are chosen:  $\lambda$ = \{0, 0.25, 0.5, 0.75, 1\}. For these channels, the unitary operations $U$ and $U'$ will only be necessary for the bit phase flip channel.
  
\subsection{Amplitude damping and phase damping channels}\label{subsec:ADchannel}

The amplitude damping channel (AD) describes the process of energy dissipation, and it can be given by following the Kraus operators \cite{Book:Krauss},
\begin{equation}\label{eq:ADKraus}
    K_0=\left( 
		\begin{array}{cc}
		1 & 0 \\ 
		0 & \sqrt{1-\lambda} \\ 
		\end{array} \right),\;\;\;\;\;\;\; K_1=\left( 
		\begin{array}{cc}
		0 & \sqrt{\lambda} \\ 
		0 & 0 \\ 
		\end{array} \right). 
\end{equation}
Table \ref{Tab:ADchannel} presents the parameters of SK-decomposition for the implementation of the AD channel. For each $\lambda$, there is a respective set of parameters given by $\gamma_1$, $\gamma_2$ and the unitary operations $U$ and $U'$ that allow us to implement the operation described previously by equations (\ref{eq:QWP+HWP+QWP-Matrix}) and (\ref{Eq:Exp-R(gamma)}). As defined in Eqs.(\ref{Eq:Exp-R(gamma)}),  $\gamma_1$ and $\gamma_2$ are related to the Dove prism operation whereas $U$ and $U'$ describe the transformations implemented by the set of wave plates (QWP-HWP-QWP). It is important to note that all these parameters are obtained by controlling the DP angle $\gamma/2$, the HWP angle $\tau$ and the QWP's $\eta_1$ and $\eta_2$. For the AD channel, the set of wave plates is not necessary.

By comparing Eqs.(\ref{eq:ADKraus}) and (\ref{eq:KrausOp}), we can see that $\alpha$ and $\beta$ are directly related to the decoherence parameter $\lambda$. In this way, for each $\lambda$, there is a given $\alpha$ and $\beta$ that reproduces the respective Kraus operator. Besides, since we already know $\alpha$ and $\beta$, and consequently $K_0$ and $K_1$, we can use Eq.(\ref{eq:GEMap_Kraus}) to obtain the Kraus operators in the Solovay-Kitaev decomposition for each decoherence parameter $\lambda$. The parameter $p$ is used to produce, if necessary, the convex combination between the two channels $\Ecal^e_a(\rho)$ and $\Ecal^e_b(\rho)$, as can be verified in equation (\ref{qechannel}). For the amplitude damping channel, this composition is not necessary, and thus we consider $p=1$ (see Table \ref{Tab:ADchannel}).

The phase damping channel (PD) is a quantum channel describing a process in which we have a damping phase without energy dissipation \cite{Book:Krauss}. The Kraus operators for this quantum channel are given by
\begin{equation}\label{eq:KrausOpPhaseDamp}
    K_0 =  \left(
\begin{array}{cc}
1 & 0\\
0 & \sqrt{1-\lambda}\\
\end{array}
\right), \;\;\;\;\;
  K_1 = \left(
\begin{array}{cc}
0 & 0\\
0 & \sqrt{\lambda}\\
\end{array}
\right).
\end{equation}
For the SK-decomposition of the PD channel, we set the same parameters used for the AD channel given in table \ref{Tab:ADchannel}. The only difference is that the PD channel does not have the $\sigma_x$ operation controlled by the $\ket{v}$ state of the transverse mode after the TBS.
\begin{table}[h!]
    \centering
\caption{Parameters used to implement the amplitude damping and the phase damping channels.}\label{Tab:ADchannel}
   \begin{tabular}{|c|c|c|c|c|c|c|c|}
\hline
\multirow{2}{*}{$\lambda$} & \multicolumn{6}{c|}{$\Ecal^e_a$}                                                  & \multirow{2}{*}{p} \\ \cline{2-7}
                           & $\alpha$ & $\beta$ & $\gamma_1$ & $\gamma_2$ & $U$ & $U'$ &                    \\ \hline
0                          & 0        & 0       & $\pi/2$     & $-\pi/2$    & none           & none            & 1                  \\ \hline
0.25                       & $\pi/6$  & 0       & $\pi/3$     & $-\pi/3$    & none           & none           & 1                  \\ \hline
0.5                        & $\pi/4$  & 0       & $\pi/4$     & $-\pi/4$    & none           & none            & 1                  \\ \hline
0.75                       & $\pi/3$  & 0       & $\pi/6$     & $-\pi/6$    & none           & none            & 1                  \\ \hline
1                          & $\pi/2$  & 0       & 0           & 0           & none           & none            & 1                  \\ \hline
\end{tabular}
 \end{table}

\subsection{Bit flip channel}\label{subsec:BFchannel}

The bit flip channel is the simplest example of a noisy channel. This channel flips the state of a qubit with probability $\lambda$ and leaves it unchanged with probability $1-\lambda$. Formally, it is described by the following set of Kraus operators:
\begin{equation}\label{eq:KrausOpBitFlip}
K_0 = \sqrt{1-\lambda} \left(
\begin{array}{cc}
1 & 0\\
0 & 1\\
\end{array}
\right), \;\;
  K_1 = \sqrt{\lambda} \left(
\begin{array}{cc}
0 & 1\\
1 & 0\\
\end{array}
\right).
\end{equation}
Table \ref{Tab:BFchannel} presents the set of parameters necessary to implement the SK-decomposition for the bit flip channel in our experimental setup. 
We can see in \ref{Tab:BFchannel} that the transformations $U$ and $U'$ are not required to implement the bit flip channel, as well as no combination of $\Ecal^e_a(\rho)$ and $\Ecal^e_b(\rho)$ is necessary ($p=1$ for all $\lambda$).
\begin{table}[h!]
\centering
\caption{Parameters used to implement the bit flip channel.}\label{Tab:BFchannel}
\begin{tabular}{|c|c|c|c|c|c|c|c|}
\hline
\multirow{2}{*}{$\lambda$} & \multicolumn{6}{c|}{$\Ecal^e_a$}                                                   & \multirow{2}{*}{p} \\ \cline{2-7}
                           & $\alpha$ & $\beta$ & $\gamma_1$ & $\gamma_2$ & $U$ & $U'$ &                    \\ \hline
0                          & 0        & 0       & $\pi/2$     & $-\pi/2$    & none           & none            & 1                  \\ \hline
0.25                       & $\pi/6$  & $\pi/6$ & $\pi/2$     & $-\pi/6$    & none           & none            & 1                  \\ \hline
0.5                        & $\pi/4$  & $\pi/4$ & $\pi/2$     & 0           & none           & none            & 1                  \\ \hline
0.75                       & $\pi/3$  & $\pi/3$ & $\pi/2$     & $\pi/6$     & none           & none            & 1                  \\ \hline
1                          & $\pi/2$  & $\pi/2$ & $\pi/2$     & $\pi/2$     & none           & none            & 1                  \\ \hline
\end{tabular}
 \end{table}

\subsection{Phase flip channel}\label{subsec:PFchannel}

The phase flip channel is a bit flip in the conjugate basis, i.e., the qubit flips its phase after interacting with the environment. This channel acts according to the following operators
\begin{equation}\label{eq:KrausOpPhaseFlip}
    K_0 = \sqrt{1-\lambda} \left(
\begin{array}{cc}
1 & 0\\
0 & 1\\
\end{array}
\right), \;\;
  K_1 = \sqrt{\lambda} \left(
\begin{array}{cc}
1 & 0\\
0 & -1\\
\end{array}
\right).
\end{equation}
Table \ref{Tab:PFchannel} presents the parameters to implement the phase flip channel (Fig.\ref{Fig:ExpSetup}).
For this channel, the composition in two channels $\Ecal^e_a(\rho)$ and $\Ecal^e_b(\rho)$ is now required and is given by
\begin{equation}\label{eq:1qbitGEMap}
    \Ecal_{PF} = p \Ecal^e_a + (1-p)\Ecal^e_b = p (M_0 \rho_i M^{\dagger}_0) + (1-p) (M_1 \rho_i M^{\dagger}_1),
\end{equation}
where the parameter $p$ ($p \in$ $[0,1]$) is responsible to combine the results obtained by the channels $\Ecal^e_a$ and $\Ecal^e_b$.
The channel $\Ecal^e_a(\rho)$ is mapped in order to implement the Kraus operator $M_0$ in the Solovay-Kitaev decomposition, described by equation (\ref{eq:GEMap_Kraus}), while the channel $\Ecal^e_b(\rho)$ is used to implement the other Kraus operator $M_1$ in this decomposition. In other words, when the parameters presented in Table \ref{Tab:PFchannel} are chosen to implement $\Ecal^e_a(\rho)$, we are considering that the experimental setup produces the operation related to the state described by the term $M_0 \rho_i M^{\dagger}_0$ in the operator sum representation for the phase flip map. On the other hand, the parameters showed in the same table used to implement $\Ecal^e_b(\rho)$ give us the operation related to the other Kraus operator, such that the transformation is described by the term $M_1 \rho_i M^{\dagger}_1$.
Therefore, it is interesting to note that in our SK-decomposition to the phase flip channel, $p$ plays the role of the decoherence parameter $\lambda$. 

To implement the phase flip channel experimentally, following the strategy of Ref.\cite{SK-Lu_Wei:17}, we first set up our experiment with the parameters of the channel $\Ecal^e_a$ and capture all images related to this first channel by varying $\lambda$. After this, we use the normalized images to reconstruct their respective density matrices. The second step is to set up the experiment with the parameters of the channel $\Ecal^e_b$, capture all images, and use the normalized intensities to reconstruct all density matrices related to this channel. Once all density matrices for $\Ecal^e_a$ and $\Ecal^e_b$ are obtained, we may choose the parameter $p$ to produce $\Ecal$ related to the evolution of the phase flip channel given by the operator sum representation. A one-way circuit to perform the phase flip channel can be obtained by duplicating the experimental setup using a second laser and mixing both outputs by controlling the relative intensities to simulate different $p$. In SK-decomposition for the phase flip channel, we also do not need to use the transformations $U$ and $U'$.
\begin{table}[h!]
  \centering
\caption{Parameters used to implement the phase flip channel.}\label{Tab:PFchannel}
\begin{tabular}{|c|c|c|c|c|c|c|c|c|c|c|c|c|c|}
\hline
\multirow{2}{*}{$\lambda$}
& \multicolumn{6}{c|}{$\Ecal^e_a$}                                                  & \multicolumn{7}{c|}{$\Ecal^e_b$}                                                         \\ \cline{2-14} 
                           & $\alpha$ & $\beta$ & $\gamma_1$ & $\gamma_2$ & $U$ & $U'$ & $\alpha$ & $\beta$ & $\gamma_1$ & $\gamma_2$ & $U$ & $U'$ & p    \\ \hline
0                          & $\pi$    & 0       & $-\frac{\pi}{2}$    & $\frac{\pi}{2}$     & none           & none            & 0        & 0       & $\frac{\pi}{2}$     & $-\frac{\pi}{2}$    & none           & none            & 0    \\ \hline
0.25                       & $\pi$    & 0       & $-\frac{\pi}{2}$    & $\frac{\pi}{2}$     & none           & none            & 0        & 0       & $\frac{\pi}{2}$     & $-\frac{\pi}{2}$    & none           & none            & 0.25 \\ \hline
0.5                        & $\pi$    & 0       & $-\frac{\pi}{2}$    & $\frac{\pi}{2}$     & none           & none            & 0        & 0       & $\frac{\pi}{2}$     & $-\frac{\pi}{2}$    & none           & none            & 0.5  \\ \hline
0.75                       & $\pi$    & 0       & $-\frac{\pi}{2}$    & $\frac{\pi}{2}$     & none           & none            & 0        & 0       & $\frac{\pi}{2}$     & $-\frac{\pi}{2}$    & none           & none            & 0.75 \\ \hline
1                          & $\pi$    & 0       & $-\frac{\pi}{2}$    & $\frac{\pi}{2}$     & none           & none            & 0        & 0       & $\frac{\pi}{2}$     & $-\frac{\pi}{2}$    & none           & none            & 1    \\ \hline
\end{tabular}
\end{table}

\subsection{Bit phase flip channel}\label{subsec:PPFchannel}

The bit phase flip channel describes a change that involves the bit flip as well as its phase. The channel acts as a follow
\begin{equation}\label{eq:KrausOpBitPhaseFlip}
    K_0 = \sqrt{1-\lambda} \left(
\begin{array}{cc}
1 & 0\\
0 & 1\\
\end{array}
\right), \;\;
  K_1 = \sqrt{\lambda} \left(
\begin{array}{cc}
0 & -i\\
i & 0\\
\end{array}
\right),
\end{equation}
Table \ref{Tab:BPFchannel} presents the parameters necessary to implement the SK-decomposition of this channel. In this case, we use the composition of the two channels $\Ecal^e_a(\rho)$ and $\Ecal^e_b(\rho)$ to construct the map $\Ecal(\rho)$. Again, the channel $\Ecal^e_a(\rho)$ is used to implement the Kraus operator $M_0$, described by equation (\ref{eq:GEMap_Kraus}), whereas channel $\Ecal^e_b(\rho)$ is necessary to implement the other Kraus operator $M_1$. Then, the map $\Ecal_{BPF}$ is constructed according to Eq.(\ref{eq:1qbitGEMap}).
\begin{figure*}[!htb]
	\centering
	\includegraphics[scale=0.45]{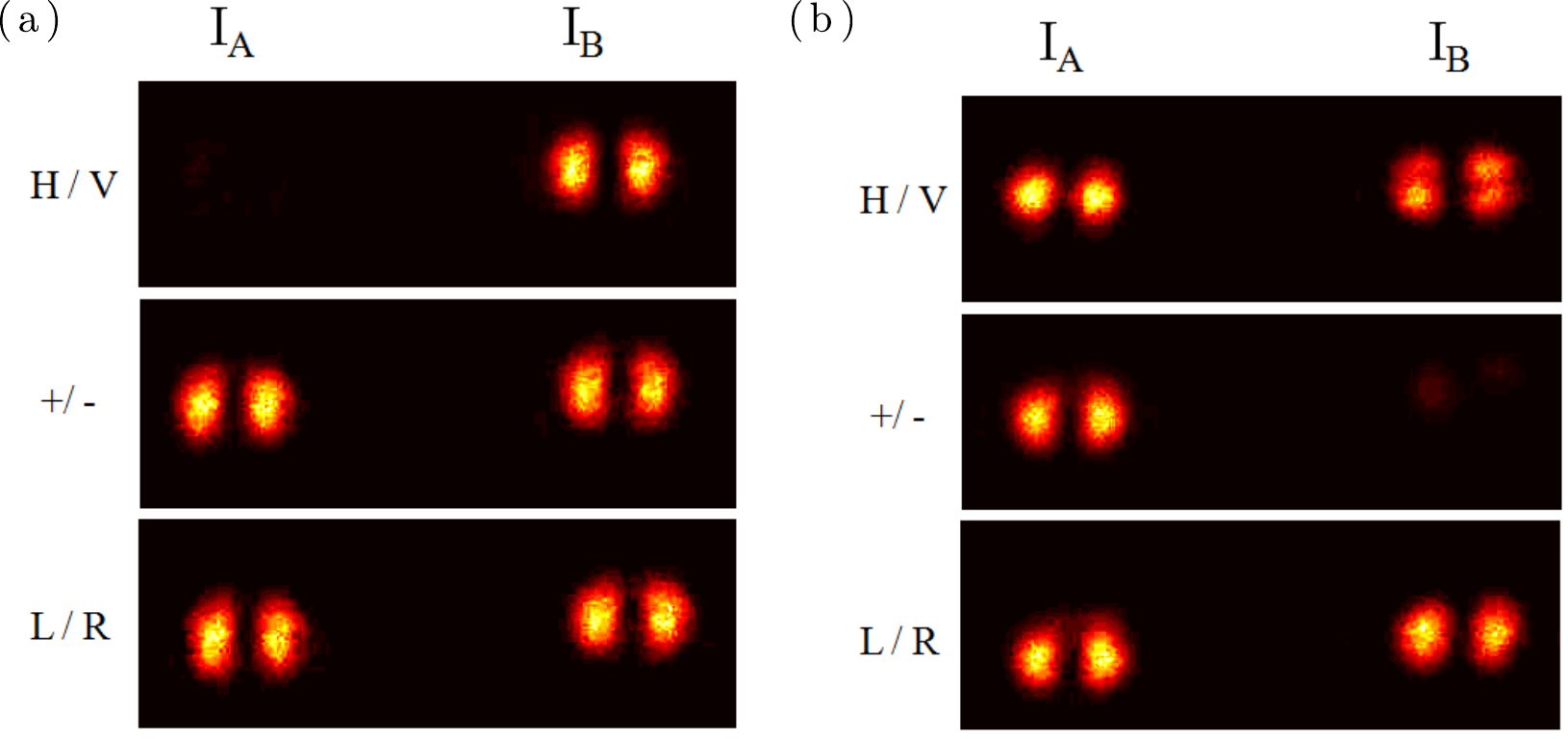}
	\caption{Hot color in tomographic images for the initial states $\ket{V}$ and $\ket{+}$, with intensities $I_{A}$ and $I_{A}$ for the basis $\{A,B\}$ = $\{H,V\}$, $\{D,AD\}$, and $\{L,R\}$. (a) Initial polarization state $\ket{\psi_i} = \ket{V}$ and (b) Initial polarization state $\ket{\psi_i} = \ket{+}$.}
	\label{fig:modes}
\end{figure*}
In this channel, using equations (\ref{eq:GEMap_Kraus}) and (\ref{eq:KrausOpBitPhaseFlip}) to map the Kraus operators, we observe that only $U$ is necessary. Hence, we label $U$ as $U_{\text{BPF}}$, and it is written as 
\begin{equation} \label{eq:U_BPF}
U_{\text{BPF}} = \text{QWP}(-\pi/2)\, \text{HWP}(\pi/2)\, \text{QWP}(0) = \left(
\begin{array}{cc}
-i & 0\\
0 & i\\
\end{array}
\right)\,.
\end{equation}
It is important to comment that, as can be inspected in the Solovay-Kitaev circuit Fig.\ref{fig:Circuit-SL}, the rotation $U_{\text{BPF}}$ is implemented before the CNOT gate. Following the same protocol employed for the phase flip channel, after obtained all the density matrix for $\Ecal^e_1$ and $\Ecal^e_2$, we can choose the parameter $p$ to construct $\Ecal$. Note that the parameter $p$ ($p \in$ $[0,1]$) used to combine the channels $\Ecal^e_1$ and $\Ecal^e_2$ play the role of decoherence parameter $\lambda$.
\begin{table}[h!]
  \centering
\caption{Parameters used to implement the bit phase flip channel.} \label{Tab:BPFchannel}
\begin{tabular}{|c|c|c|c|c|c|c|c|c|c|c|c|c|c|}
\hline
\multirow{2}{*}{$\lambda$} & \multicolumn{6}{c|}{$\Ecal^e_a$}                                                           & \multicolumn{7}{c|}{$\Ecal^e_b$}                                                                   \\ \cline{2-14} 
                           & $\alpha$ & $\beta$ & $\gamma_1$     & $\gamma_2$      & $U$ & $U'$ & $\alpha$         & $\beta$ & $\gamma_1$ & $\gamma_2$ & $U$   & $U'$ & p    \\ \hline
0                          & 0        & 0       & $\frac{\pi}{2}$ & -$\frac{\pi}{2}$ & none           & none            & $\frac{\pi}{2}$ & $\frac{\pi}{2}$       & $\frac{\pi}{2}$      & $\frac{\pi}{2}$      & $U_{\text{BPF}}$ & none            & 0    \\ \hline
0.25                       & 0        & 0       & $\frac{\pi}{2}$ & -$\frac{\pi}{2}$ & none           & none            & $\frac{\pi}{2}$ & $\frac{\pi}{2}$       & $\frac{\pi}{2}$      & $\frac{\pi}{2}$       & $U_{\text{BPF}}$ & none            & 0.25 \\ \hline
0.5                        & 0        & 0       & $\frac{\pi}{2}$ & -$\frac{\pi}{2}$ & none           & none            & $\frac{\pi}{2}$ & $\frac{\pi}{2}$       & $\frac{\pi}{2}$      & $\frac{\pi}{2}$      & $U_{\text{BPF}}$ & none            & 0.5  \\ \hline
0.75                       & 0        & 0       & $\frac{\pi}{2}$ & -$\frac{\pi}{2}$ & none           & none            & $\frac{\pi}{2}$ & $\frac{\pi}{2}$       & $\frac{\pi}{2}$      & $\frac{\pi}{2}$       & $U_{\text{BPF}}$ & none            & 0.75 \\ \hline
1                          & 0        & 0       & $\frac{\pi}{2}$ & $\frac{\pi}{2}$ & none           & none            & $\frac{\pi}{2}$ & $\frac{\pi}{2}$       & $\frac{\pi}{2}$      & $\frac{\pi}{2}$       & $U_{\text{BPF}}$ & none            & 1    \\ \hline
\end{tabular}
\end{table}

\section{Results and Discussion} \label{sec:Results}

Let us start by presenting the characterization of the initial states utilized in the experiment. Depending on the SK-decomposition, we studied the evolution of the state $\ket{V}\equiv \ket{1}$ ($\phi=45^\circ$ in Eq.(\ref{eq:InitialSatate})) or the evolution of the superposition state $\ket{+}\equiv(\ket{0}+\ket{1})/\sqrt{2}$ ($\phi=22.5^\circ$ in Eq.(\ref{eq:InitialSatate})). The output intensities of the tomographic measurements for the initial states $\ket{V}$ and $\ket{+}$ are shown in Figs.\ref{fig:modes} (a)-(b), respectively. Each image is a single frame, and the relative intensities were obtained from the integration of the gray level intensities to apply Eq.(\ref{eq:Prob_Int}). Note that the tomography is only performed in the polarized state. 

\begin{figure}[!htb]
\centering
\includegraphics[scale=0.4]{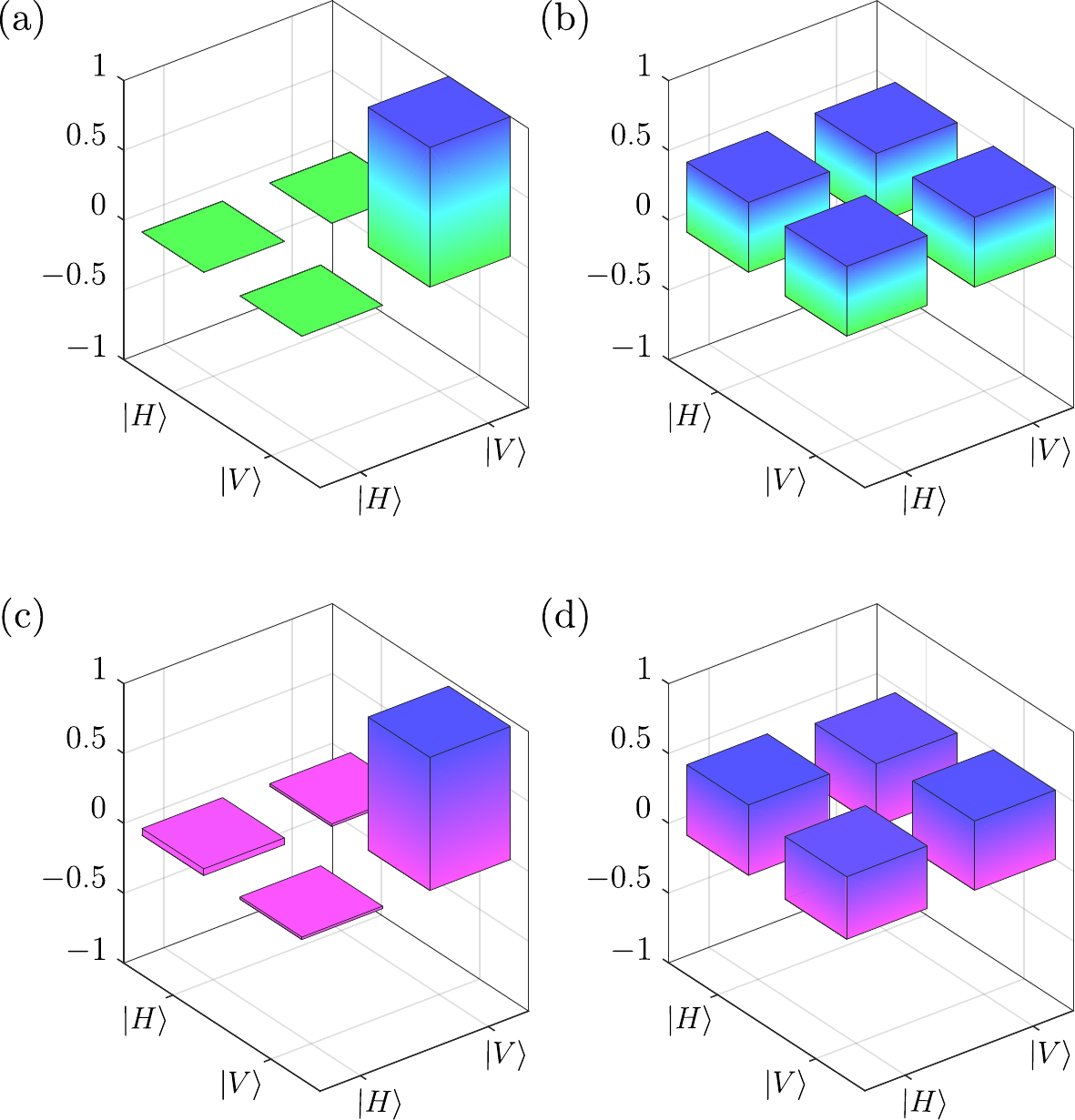}
\caption{Matrix reconstructions for initial states $\ket{V}$ and $\ket{+}$. Theoretical results in the top graphs and experimental results at the bottom. (a) Initial polarization state $\ket{\psi_i} = \ket{V}$, (b) Initial polarization state $\ket{\psi_i} = \ket{+}$, (c) Initial polarization state $\ket{\psi_i} = \ket{V}$, and (d) Initial polarization state $\ket{\psi_i} = \ket{+}$.}
\label{fig:rec_initial}
\end{figure}

The theoretical and experimental density matrix reconstruction for the initial states $\ket{V}$ ($\rho_{V}$), and $\ket{+}$ ($\rho_{+}$) is presented in Fig.\ref{fig:rec_initial}. The fidelity obtained with the experiment for the initial state, $\ket{V}$ is F = 0.9996$\pm 0.002$ and for $\ket{+}$ it is F = 0.9873$\pm 0.0217$, showing very good agreement between the theory and the experiment. 

\begin{figure*}[!htb]
 \begin{center}
 \includegraphics[scale=0.4]{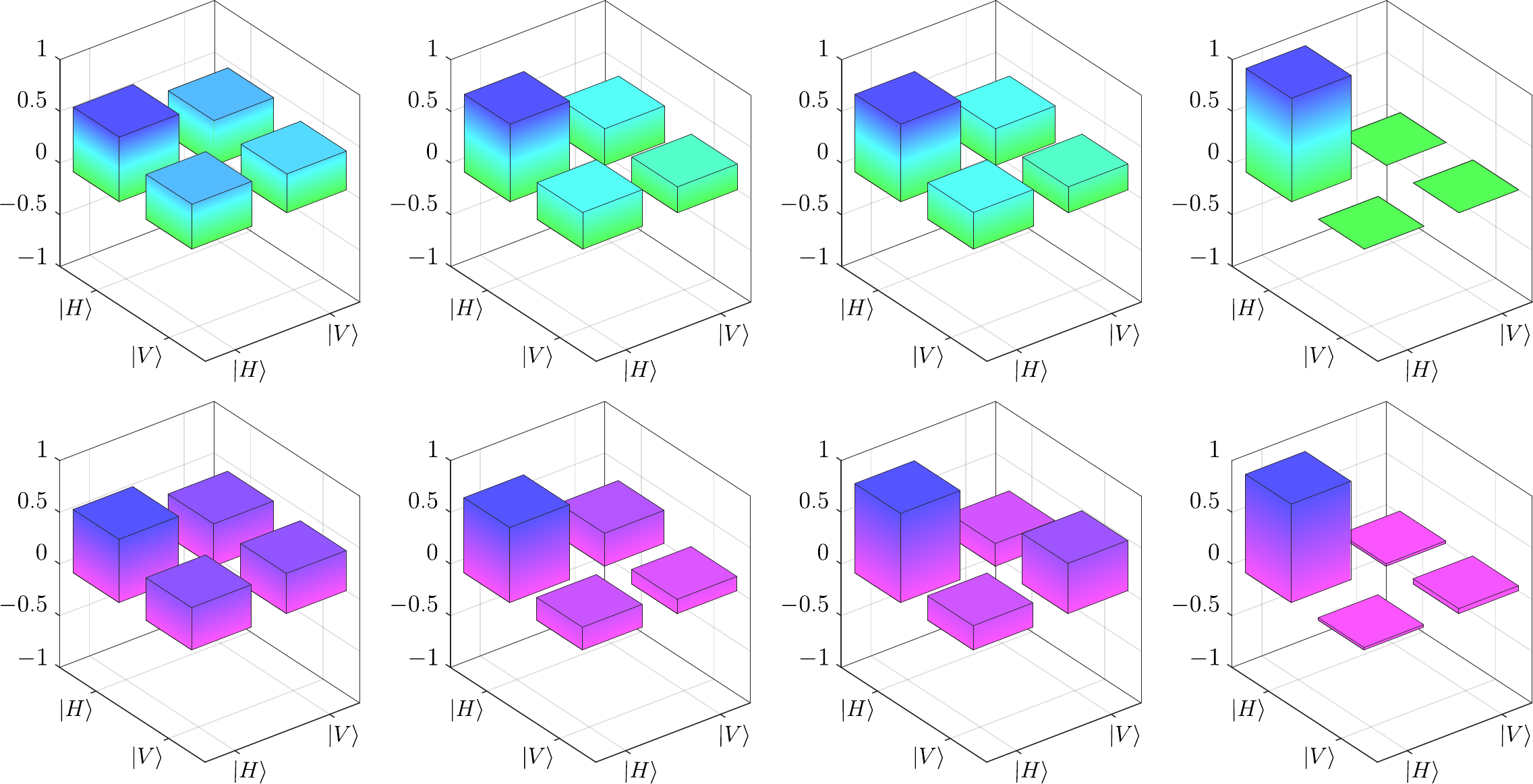}R
\end{center}
\caption{Theoretical (green and blue color) and experimental (purple and blue color) matrix elements for the amplitude damping channel reconstruction with the initial state $\ket{+}$. From left to right the parameter $\lambda$ was set as \{0.25,0.5,0.75,1\}.}
 \label{Reconstruction_AmplitudeDamp_+}
\end{figure*}

\subsection{Amplitude damping and phase damping channels}\label{subsec:Res-ADchannel}

The density matrix reconstruction for the initial polarization states $\ket{+}$ is shown in Fig.\ref{Reconstruction_AmplitudeDamp_+}. The results corresponds to $\lambda =0.25, 0.5, 0.75, 1$, from left to right. The theoretical prediction is shown in the top of Fig.\ref{Reconstruction_AmplitudeDamp_+} (green and blue colors). A gradual decay to the ground state is expected for the AD channel with the initial state $\ket{+}$ going to $\ket{H}$. The density matrix obtained experimentally is presented at the bottom of Fig.\ref{Reconstruction_AmplitudeDamp_+} (purple and blue colors). 
\begin{figure}[!htb]
 \begin{center}
 \includegraphics[scale=0.7]{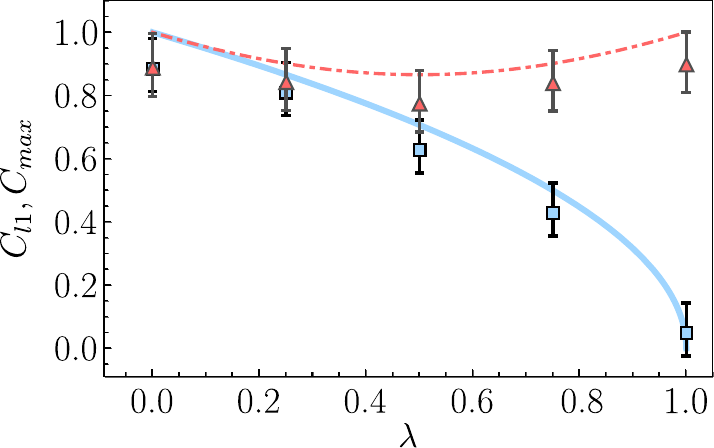}
\end{center}
\caption{Theoretical (solid line - blue online) and experimental (squares - blue online) for $l_1$-norm coherence and theoretical (dashed-line - red online) and experimental (triangles - red online) for maximal coherence as a function of the decoherence parameter for the AD channel acting on the initial state $\ket{+}\,$.}
 \label{Fig:Res-AD_+}
\end{figure}

The behaviour of the $l_1$-norm coherence $C_{l1}(\lambda)$ and Maximal coherence $C_{max}(\lambda)$ as functions of the decorehence parameter $\lambda$ for the initial state $\ket{+}$ ($r_x = 1$ and $r_y = 0$) is presented in Fig.\ref{Fig:Res-AD_+}. The experimental results of the $C_{l1}(\lambda)$ are represented by squares and the solid line (blue color) is the theoretical prediction. The experimental results obtained for the maximal coherence $C_{max}(\lambda)$ are represented by triangles while the theoretical prediction is represented by dot-solid line (red color).

\begin{figure}[!htb]
	\centering
	\includegraphics[scale=0.4]{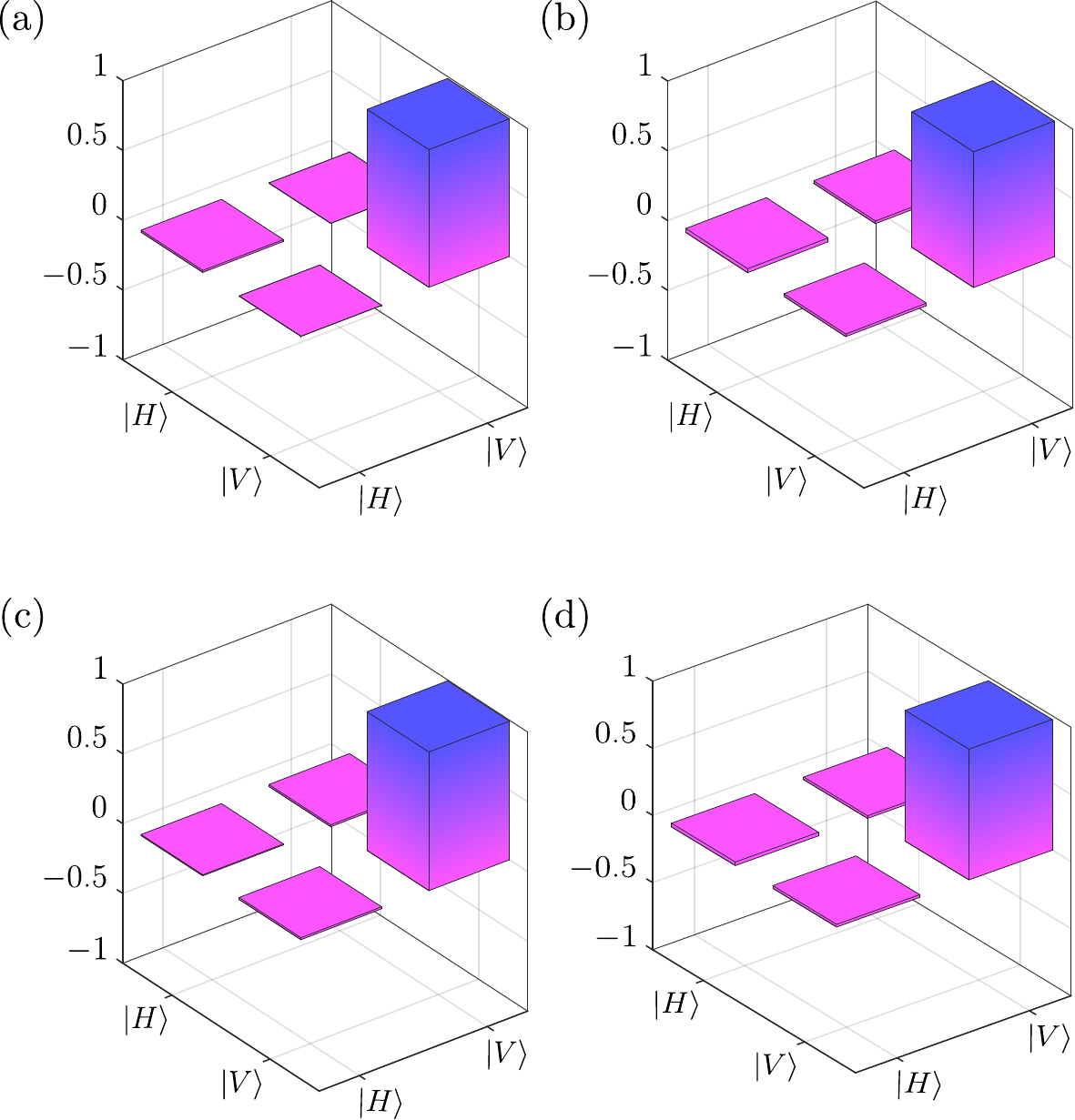}
	\caption{Experimental density matrix reconstruction for the initial state $\ket{V}$ under the action of the phase damping channel (PD). (a) $\lambda = 0.25$: Fidelity with respect to the theoretical predictions is F = $0.9861 \pm 0.0216$, (b) $\lambda = 0.5$: Fidelity with respect to the theoretical predictions is F = $0.9721 \pm 0.0220$, (c) $\lambda = 0.75$: Fidelity with respect to the theoretical predictions is F = $0.9933 \pm 0.0020$, and (d) $\lambda = 1.0$: Fidelity with respect to the theoretical predictions is F = $0.9732 \pm 0.0220$. }
	\label{Reconstruction_PhaseDamp_V}
\end{figure}

The SK-decomposition of the PD channel (Fig.\ref{Reconstruction_PhaseDamp_V}) shows the density matrix reconstruction for $\lambda~>~0$ considering the initial state $\ket{V}$. The reconstructions present a high fidelity. Nevertheless, it is worth to mention that the density matrix population does not change during the evolution. 

\begin{figure}[!htb]
 \begin{center}
 \includegraphics[scale=0.7]{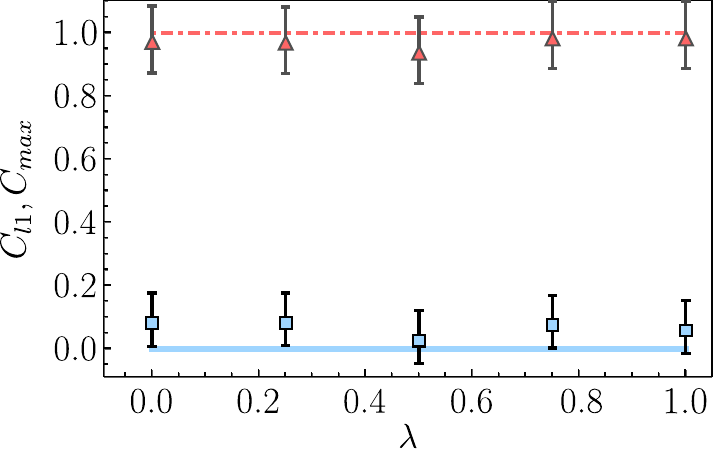}
\end{center}
\caption{Theoretical (solid line - blue online) and experimental (squares - blue online) for $l_1-$norm coherence and theoretical (dot-dashed line - red online) and experimental (triangles - red online) for maximal coherence as a function of the decoherence parameter for the $\ket{V}$ state under the action of the PD.}
 \label{Fig:Res-PD_V}
\end{figure}

Figure \ref{Fig:Res-PD_V} presents the coherence $C_{l1}(\lambda)$  (squares and solid line - blue online) and maximal coherence $C_{max}(\lambda)$  (triangles and dot-dashed line - red online) as functions of the decorehence parameter $\lambda$ for the initial state $\ket{V}$ under the action of the phase damping channel (PD). Note that the analysis of the PD channel is considered with respect to the decoherence parameter $\lambda\in[0,1]\,$.

As expected, the $l_1-$norm coherence shows a freeze behavior in the minimum value ($C(\lambda)=0$) since neither the initial state $\ket{V}$ ($r_x = r_y = 0$) nor any other obtained in this evolution present a coherent superposition. The maximal coherence also exhibits the freezing behavior, but now this happens in its maximum value $C_{max}(\lambda)=1$. This can be understood by the fact that the phase damping does not change the purity of the quantum state. Although the parametrization time is slightly different, our experimental results for these two classes of coherence are in excellent agreement with what was obtained in Ref.\cite{Passos_Markov} using another experimental setup.

\subsection{Bit flip channel}\label{subsec:Res-BFchannel}

\begin{figure}[!htb]
	\centering
	\includegraphics[scale=0.4]{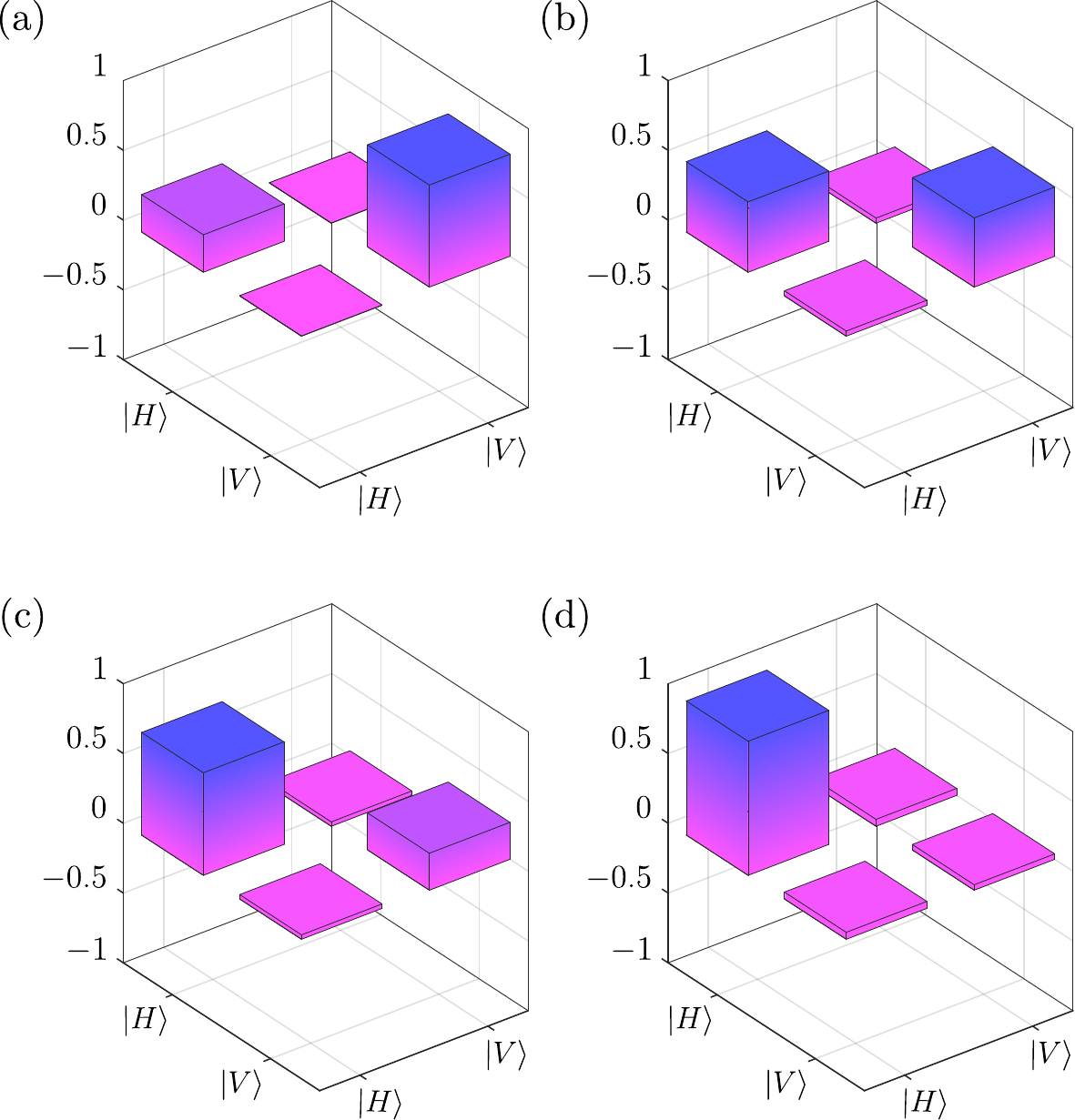}
	\caption{Experimental matrix elements reconstructions of the initial state $\ket{V}$ for the bit flip channel (BF). (a) $\lambda = 0.25$: Fidelity with respect to the theoretical predictions is F = $0.9945 \pm 0.0020$, (b) $\lambda = 0.5$: Fidelity with respect to the theoretical predictions is F = $0.9998 \pm 0.0020$, (c) $\lambda = 0.75$: Fidelity with respect to the theoretical predictions is F = $0.9996 \pm 0.0020$, and (d) $\lambda = 1.0$: Fidelity with respect to the theoretical predictions is F = $0.9609 \pm 0.0317$.}
	\label{Reconstruction_BitFlip_V}
\end{figure}

For the bit flip (BF) SK-decomposition, Fig.\ref{Reconstruction_BitFlip_V} shows the density matrix reconstruction for each $\lambda$ for the initial polarization state $\ket{V}$ with the respective fidelity F. Observe that the bit-flip occurs for $\lambda = 1$. 

\begin{figure}[!htb]
 \begin{center}
 \includegraphics[scale=0.7]{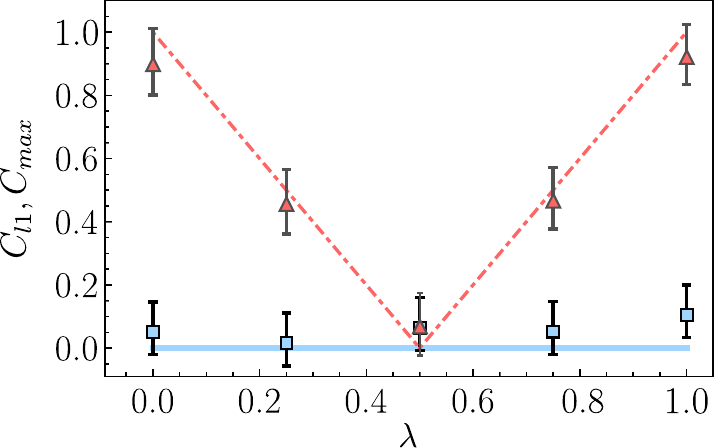}
\end{center}
\caption{Theoretical (solid line - blue online) and experimental (squares - blue online) results for $l_1-$norm coherence and theoretical (dot-dashed - red online) and experimental (triangles - red online) for maximal coherence as a function of the decoherence parameter for the BF channel prepared in the $\ket{V}$ state.}
 \label{Fig:Res-BF_V}
\end{figure}

Concerning the coherence behavior of the state for the BF SK-decomposition, Fig.\ref{Fig:Res-BF_V} presents the  $l_1$-norm coherence $C_{l1}(\lambda)$ (experiment: blue squares, theory: blue solid line) and maximal coherence $C_{max}(\lambda)$ (experiment: red triangles, theory: red dashed-dot line) as functions of the decorehence parameter $\lambda$ for the initial polarization state  $\ket{V}$. We can observe that the $l_1-$norm coherence shows a freezing behavior in the minimum value of $C(\lambda) = 0$. This happens because we started the evolution with state $\ket{V}$ that has no coherence in the H-V basis, and, as expected, it is not possible to obtain any coherent superposition in this evolution. On the other hand, the maximal coherence ($C_{max}(\lambda)$) varies with the parameter $\lambda$. The polarization state $\ket{V}$ is pure and consequently its maximum coherence manifests at the maximum value ($C_{max}(0) = 1$). However, for $p=0.5$ we obtain a maximal mixture between the states $\ket{H}$ and $\ket{V}$ and the maximal coherence exhibit the minimum value $C_{max}(0.5) = 0$). In this case, the qubit has 50$\%$ probability for flipping its state.

\subsection{Phase flip channel}\label{subsec:Res-PFchannel2}

\begin{figure}[!htb]
	\centering
	\includegraphics[scale=0.4]{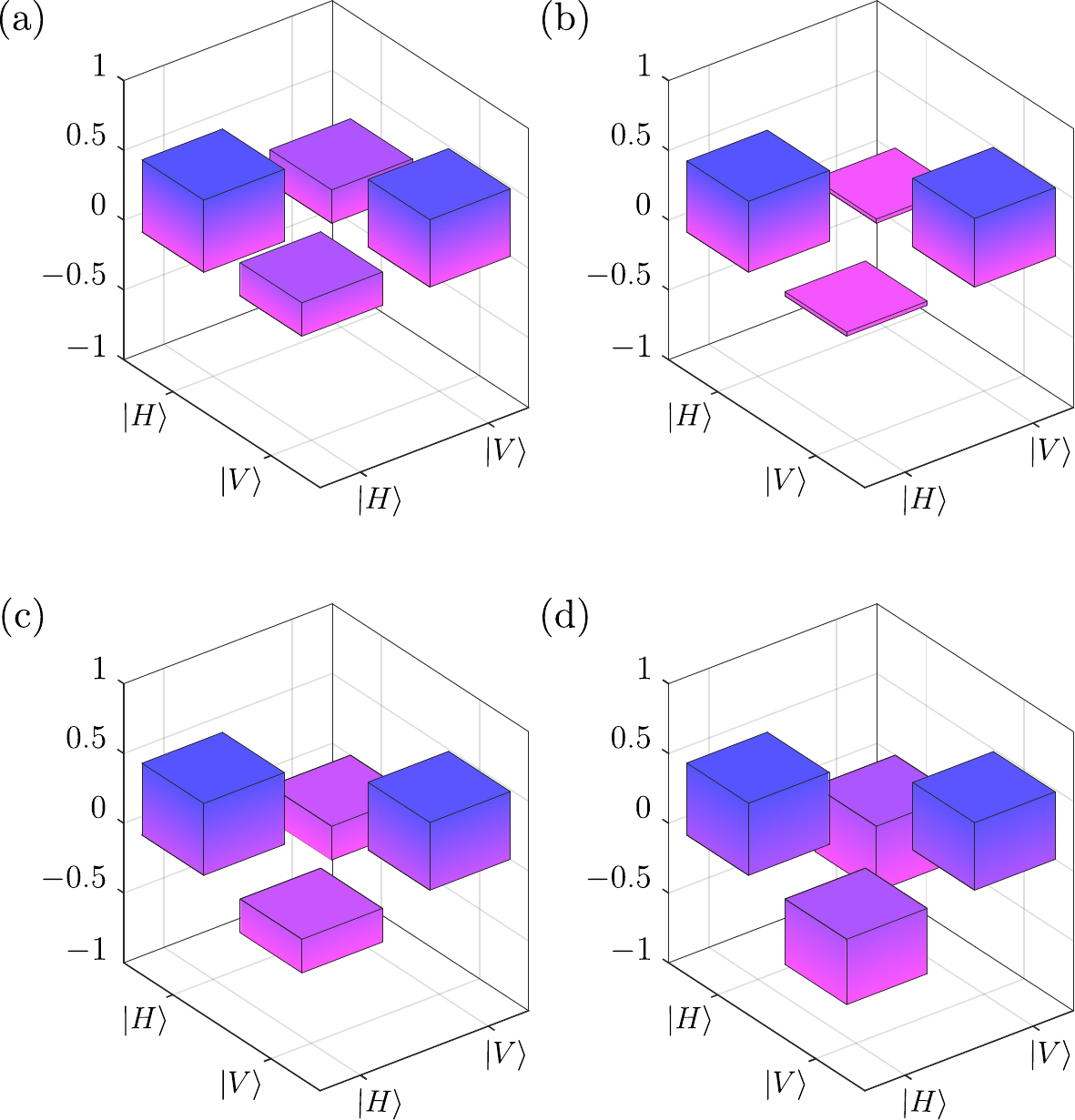}
	\caption{Experimental matrix reconstructions of the initial state $\ket{+}$ for the phase flip channel (PF). (a) $\lambda = 0.25$: Fidelity with respect to the theoretical predictions is F = $0.9996 \pm 0.0020$, (b) $\lambda = 0.5$: Fidelity with respect to the theoretical predictions is F = $0.9998 \pm 0.0020$, (c) $\lambda = 0.5$: Fidelity with respect to the theoretical predictions is F = $0.9998 \pm 0.0020$, and (d) $\lambda = 1.0$: Fidelity with respect to the theoretical predictions is F = $0.9996 \pm 0.0020$.}
	\label{Reconstruction_PhaseFlip_+}
\end{figure}

The SK-decomposition of the phase flip (PF) channel for the density matrix reconstruction is presented in Fig.\ref{Reconstruction_PhaseFlip_+}. The analysis was done for the initial state $\ket{+}$ (Fig.\ref{fig:rec_initial}). We can observe a high fidelity for each $\lambda$, indicating that we achieved the SK-decomposition successfully, since the final state is $ \ket{-}$.

\begin{figure}[!htb]
 \begin{center}
 \includegraphics[scale=0.7]{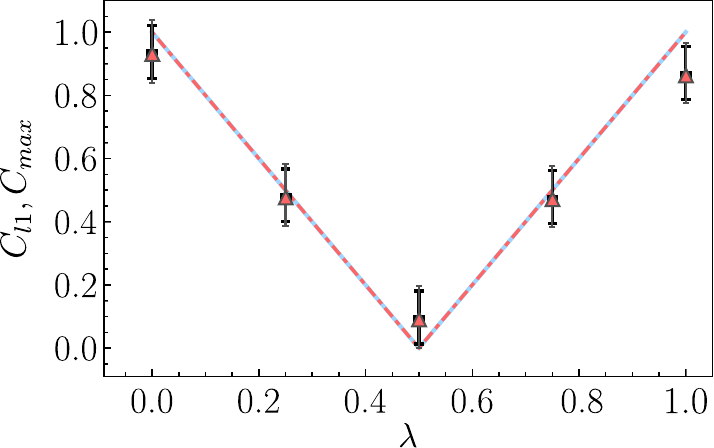}
\end{center}
\caption{Theoretical (solid - blue online) and experimental (squares - blue online) for $l_1-$norm coherence and theoretical (dot-dashed - red online) and experimental (triangles - red online) for maximal coherence as a function of the decoherence parameter for the PF channel prepared in the $\ket{+}$ state.}
 \label{Fig:Res-PF_+}
\end{figure}

Considering as the initial state a maximally coherent polarization state $\ket{+}$, Figure \ref{Fig:Res-PF_+} presents the evolution of the $l_1-$ norm coherence $C(\lambda)$ (squares, solid line - blue online) and maximal coherence $C_{max}(\lambda)$ (triangles, dot-dashed line - red online) as a function of the decoherence parameter $\lambda$. It is possible to observe that, in this case, we do not have a freezing behavior for $C(\lambda)$ and $C_{max}(\lambda)$. As we started with a maximally coherent state, the coherence $C(\lambda)$ has a maximum value. Throughout the evolution, the initial superposition is lost, and consequently, $C(\lambda)$ presents a minimum value ($C=0$) at $\lambda = 0.5$. For $p=1$, the channel has already flipped the relative phase between the polarization components $\ket{H}$ and $\ket{V}$ of the initial state, yielding another maximally coherent state $\ket{-}$ which also has a maximum value for the coherence $C(\lambda)$. The maximal coherence $C_{max}(\lambda)$ shows a similar behavior since we started the evolution with a pure state $\ket{+}$, where $C_{max}(0)=1$. For $\lambda = 0.5$, we have a maximally mixed state and, consequently, we obtain $C_{max}(0.5)=0$. As discussed above, for $\lambda = 1$, the state is $\ket{-}$, and consequently, $C_{max}$ exhibits a maximum value.

\subsection{Bit phase flip channel}\label{subsec:Res-PFchannel}

\begin{figure}[!htb]
	\centering
	\includegraphics[scale=0.4]{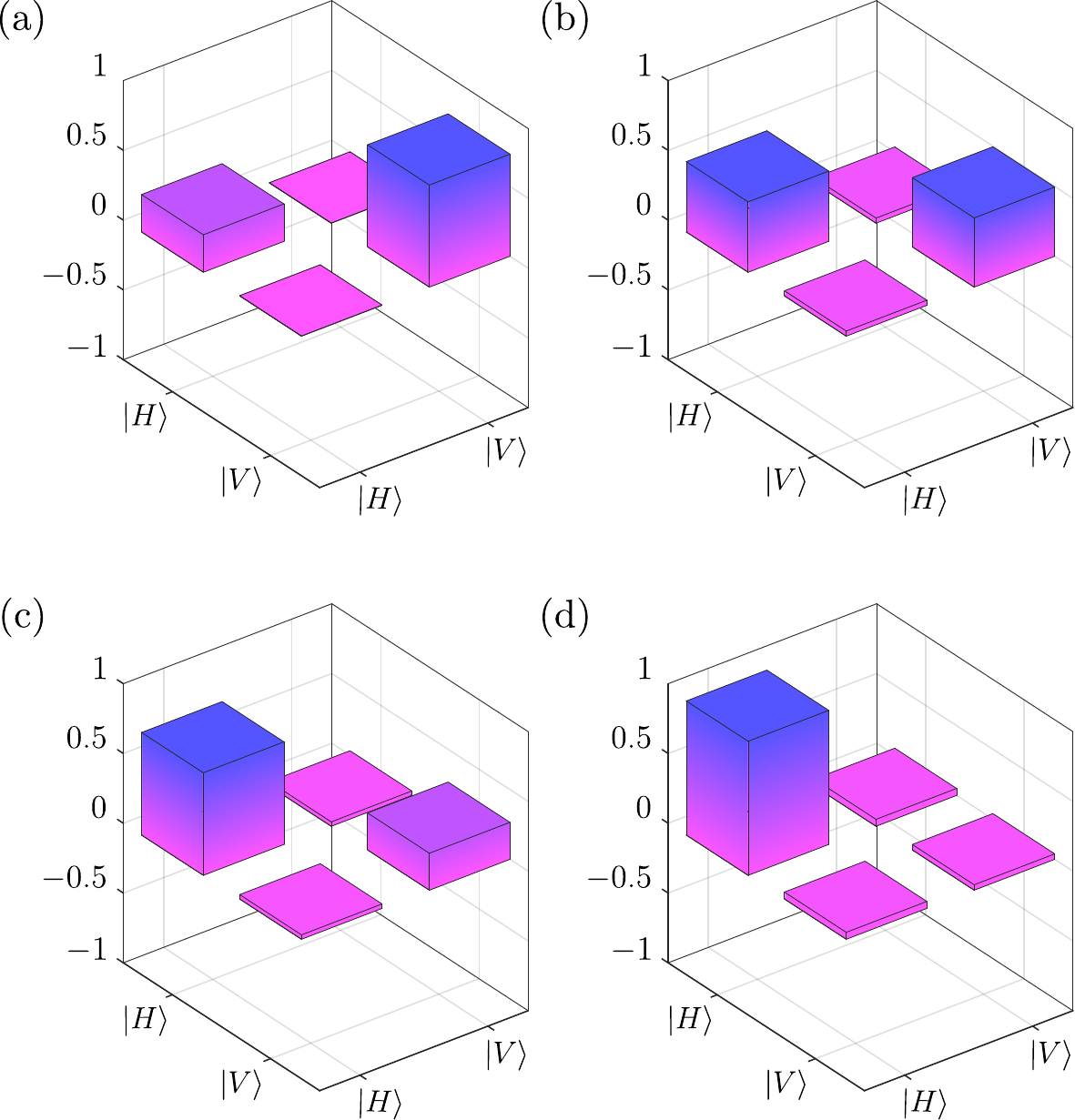}
	\caption{Experimental matrix elements reconstructions of the initial state $\ket{+}$ for the Bit phase flip channel (BPF). (a) $\lambda = 0.25$: Fidelity with respect to the theoretical predictions is F = $0.9996 \pm 0.0020$, (b) $\lambda = 0.5$: Fidelity with respect to the theoretical predictions is F = $0.9994 \pm 0.0020$, (c) $\lambda = 0.75$: Fidelity with respect to the theoretical predictions is F = $0.9996 \pm 0.0020$, (d) $\lambda = 1.0$: Fidelity with respect to the theoretical predictions is F = $0.9998 \pm 0.0020$.}
	\label{Reconstruction_BitPhaseFlip_+}
\end{figure}

The last channel decomposed in this work is the bit phase flip. The density matrix reconstruction for this channel is shown in Fig.\ref{Reconstruction_BitPhaseFlip_+}, for the initial polarization state $\ket{+}$, and final state $\ket{-}$.

\begin{figure}[!htb]
 \begin{center}
 \includegraphics[scale=0.7]{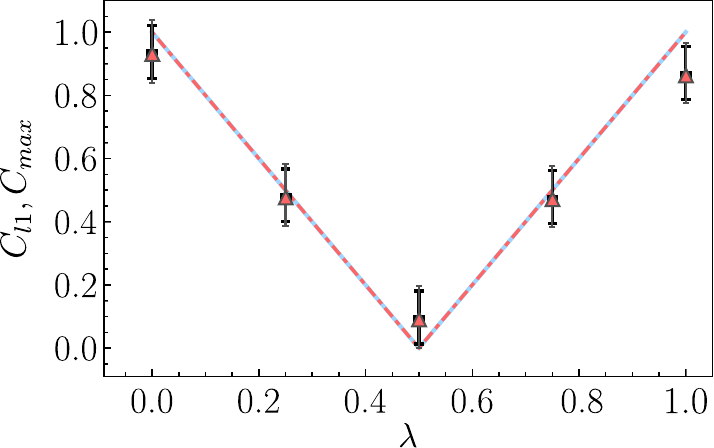}
\end{center}
\caption{Theoretical (solid - blue online) and experimental (squares - blue online) for $l_1-$norm coherence and theoretical (dot-dashed - red online) and experimental (triangles - red online) for maximal coherence as a function of the decoherence parameter for the state $\ket{+}$ undergoing the BPF channel.}
 \label{Fig:Res-BPF_+}
\end{figure}

The evolution of $l_1-$norm coherence $C_{l1}(\lambda)$ (squares, solid line - blue online) and maximal coherence $C_{max}(\lambda)$ (triangles, dot-dashed line, red online) for bit phase flip channel is shown in Fig.\ref{Fig:Res-BPF_+} considering a maximally coherent polarization state $\ket{+}$ as the initial state. As can be seen, we do not have any freezing behavior for $C_{l1}(\lambda)$ and $C_{max}(\lambda)$. Analyzing $C_{l1}(\lambda)$, we observe the expected evolution. Since the initial state is maximally coherent, we have that $C_{l1}(0)=1$. On the other hand, according to the system evolution controlled by $\lambda$ values, we observe that the coherent superposition is lost. When $\lambda=0.5$ the state does not present any coherent superposition and, consequently, we obtain $C_{l1}(0.5)=0$. However, for $\lambda = 1$, the channel has already flipped both the bit and relative phase, yielding the maximally coherent state $\ket{-}$. This final state produced by the evolution in this channel has the maximum value of $C_{l1}(\lambda=1) = 1$.

Maximal coherence $C_{max}$  starts in maximum value since the initial state is a pure state. On the other hand, when $\lambda = 0.5$ we have a maximally mixed state and consequently $C_{max}(0.5) = 0$. As the state produced by the evolution in this channel is also a maximally pure state ($\ket{-}$), the maximal coherence returns to the maximal value $C_{max}(1) = 1$. 

Finally, it is important to comment that the small differences between all theoretical and experimental results come from the limited visibility of the interferometers and the intensity sensitivity of the CCD camera.

\section{Conclusions}\label{sec:conclusions}

In conclusion, we have investigated the Solovay-Kitaev decomposition of single qubit quantum channels using the spin-orbit modes of a laser beam. The implementation of arbitrary quantum channels on the polarization (spin) degree of freedom was achieved using the transverse mode structure (orbit) as the ancillary qubit. This allowed the easy realization of the local unitary operations and controlled gates needed for implementing the required channels. Moreover, our approach with an intense laser source gave us direct access to the noisy channels' statistical properties without resorting to single-photon sampling. Both density matrix reconstruction and coherence analysis of the qubit in the channel show the decomposition's success through an excellent agreement between the theoretical predictions and the experimental results. It is important to mention that the  encoding  of  two  qubits  on  a  single  light  beam  constitutes  an  enormous  advantage  for  quantum  information processing,  when nonlocality is not required.  In general,  entanglement is vulnerable to decoherence because the entangled  systems  are  usually  subejct  to  independent  noise  sources  when  spatially  separated.   This  difficulty  is unavoidable in quantum information protocols that rely on nonlocal correlations.  When this is not the case,  as in  the  quantum  key  distribution  protocol  without  a  shared  reference  frame  \cite{PRA.77.Cadu-Cripto,marrucci:2012}, it is preferable that the two qubits are encoded on the same physical object, since they would experience the same noise source. For example, distant photon pairs experience independent random phase changes due to atmospheric turbulence and an initially entangled state is inevitably driven to a separable mixed state \cite{Gopaul:2007,Ge:2015}. This difficulty does not apply to spin-orbit entangled modes since both degrees of freedom experience the same turbulent fluctuations and the same random phase changes. This is why spin-orbit modes are candidates for robust optical communication in free space \cite{Lochab:2017}. 
This robustness explains the high fidelity of more than 98$\%$ obtained in our experiment, showing that decoherence and dissipative effects in general are overruled.
For the spin-orbit encoding of two qubits on single photons, the source of decoherence comes mainly from a small imprecision in measuring the polarization and orbital degrees of freedom.
We believe that the present architecture has an enormous potential as a platform for investigations in quantum thermodynamics, and for addressing to fundamental properties of channel capacities, where either controllable reservoirs or channels are required. Advancements along those lines shall be presented elsewhere.

\section*{Acknowledgments}
The Authors acknowledge financial support from the Brazilian funding agencies Conselho Nacional de Desenvolvimento Cient\'{\i}fico e Tecnol\'ogico (CNPq), 
Funda\c{c}\~ao Carlos Chagas Filho de Amparo \`a Pesquisa do Estado do Rio de Janeiro (FAPERJ), 
Coordena\c{c}\~ao de Aperfei\c{c}oamento de Pessoal de N\'{\i}vel Superior (CAPES) (Finance Code 001), 
and the Brazilian National Institute for Science and Technology of Quantum Information (INCT-IQ). AOJ acknowledges financial support by the Foundation for Polish Science through TEAM-NET project (contract no. POIR.04.04.00-00-17C1/18-00).

\appendix

\section{The transverse mode beam splitter}\label{Ap:MZIM+HWP}

\begin{figure}[!htb]
 \begin{center}
 \includegraphics[scale=0.7,clip,trim=0mm 0mm 0mm 0mm]{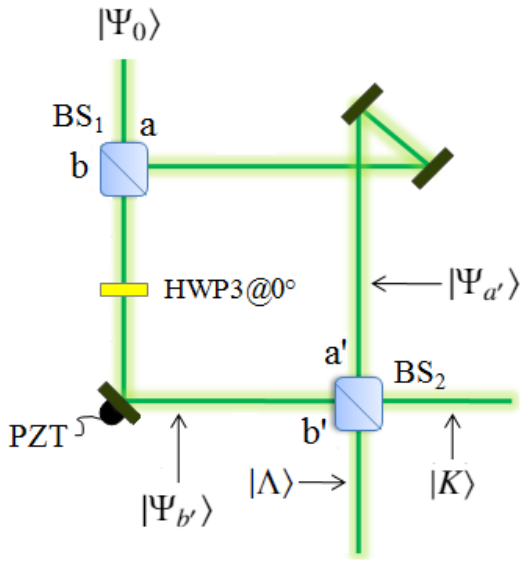} 
\end{center}
\caption{Experimental circuit to implement the new configuration of the MZIM. This device will allow us to produce projective measurements in the first-order transverse modes.}
 \label{fig:MZIM-HWP}
\end{figure}

 In this appendix we will discuss the action of the HWP inserted in the MZIM. Fig.\ref{fig:MZIM-HWP} presents the apparatus in detail, marking the spin-orbit state in each part of this optical circuit. 
The principle of the MZIM is directly related to the action of mirror reflections on polarization and transverse modes. When a laser beam is reflected by a vertical mirror (horizontal plane of incidence), the vertical polarization is not affected ($\ket{V}\rightarrow\ket{V}$) while the horizontal polarization is inverted and acquires a minus sign ($\ket{H}\rightarrow -\ket{H}$). Analogously, a first order Hermite-Gaussian mode is affected in the same way $\ket{v}\rightarrow\ket{v}$ and $\ket{h}\rightarrow -\ket{h}$. Therefore, using the definition given in Eq.\eqref{eq:QWP+HWP+QWP-Matrix2}, the spin-orbit transformation performed by each mirror reflection is represented by $H_0\otimes H_0\,$. Moreover, a half-wave plate affects only the polarization part of the spin-orbit state and can be represented by $H_0\otimes I\,$. Now we can take into account all transformation steps inside the modified MZIM to understand how it works.

Let us consider a general spin-orbit mode given by 
\begin{equation}\label{eq:Ap-psi}
    \ket{\Psi_0} = \frac{\ket{\varphi_1}\otimes\ket{h} + \ket{\varphi_2}\otimes\ket{v}}{\sqrt{2}}\,,
\end{equation}
where $\ket{\varphi_{1,2}}$ are arbitrary polarization states. This mode enters the interferometer through port $a$ of the 50/50 beam splitter BS1, follows two paths with different transformation sequences and 
arrives at a second 50/50 beam splitter BS2, where it exits the interferometer through two output ports. 
The input-output relations for both beam splitters are given by the 2x2 unitary matrix 
\begin{equation}\label{eq:Ap-BS1}
  \text{BS} \equiv \frac{1}{\sqrt{2}}
\left(
\begin{array}{cc}
1 & -1\\
1 & 1\\
\end{array}
\right)\,.
\end{equation}
Note that port $b$ of BS1 is empty. Therefore, we can compute the state evolution inside the interferometer follwing the transformations implemented in each arm:
\begin{itemize}
    \item \textbf{Transmission through BS1:} This beam passes through HWP3 oriented at 0$^\circ$, follows one reflection at the PZT-mounted mirror and reaches port $b^\prime$ of BS2 in state
    \begin{eqnarray}
        \ket{\Psi_{b^\prime}} &=& \frac{e^{i\Delta}}{2} (H_0\otimes H_0)\,(H_0\otimes I)\ket{\Psi_0} 
        \nonumber\\
        &=& \frac{e^{i\Delta}}{2} (\ket{\varphi_1}\otimes\ket{h} - \ket{\varphi_2}\otimes\ket{v})\,,
    \end{eqnarray}
    where $\Delta$ is the phase shift introduced by the PZT and $H_0^2 = I$ has been used.
    \item \textbf{Reflection at BS1:} This beam follows two reflections at the arm mirrors and reaches port 
    $a^\prime$ of BS2 in state
    \begin{eqnarray}
        \ket{\Psi_{a^\prime}} &=& \frac{1}{2} (H_0\otimes H_0)^2\ket{\Psi_0} 
        \nonumber\\
        &=& \frac{1}{2} (\ket{\varphi_1}\otimes\ket{h} + \ket{\varphi_2}\otimes\ket{v})\,.
    \end{eqnarray}
\end{itemize}
Finally, using the input-output relations of BS2 and assuming that the interferometer is balanced ($\Delta=0$), 
we find the following output states
\begin{eqnarray}
    \ket{K} &=& \frac{\ket{\Psi_{a^\prime}} + \ket{\Psi_{b^\prime}}}{\sqrt{2}} = \frac{1}{\sqrt{2}}\ket{\varphi_1}\otimes\ket{h}\;,
    \nonumber\\
    \ket{\Lambda} &=& \frac{\ket{\Psi_{a^\prime}} - \ket{\Psi_{b^\prime}}}{\sqrt{2}} = \frac{1}{\sqrt{2}}\ket{\varphi_2}\otimes\ket{v}\;.
\end{eqnarray}
Therefore, each output of the modified MZIM performs a transverse mode projection without affecting the 
polarization state attached to the projected mode. This resumes our demonstration of the 
transverse mode measurement device.


\begin{thebibliography}{61}%
\makeatletter
\providecommand \@ifxundefined [1]{%
 \@ifx{#1\undefined}
}%
\providecommand \@ifnum [1]{%
 \ifnum #1\expandafter \@firstoftwo
 \else \expandafter \@secondoftwo
 \fi
}%
\providecommand \@ifx [1]{%
 \ifx #1\expandafter \@firstoftwo
 \else \expandafter \@secondoftwo
 \fi
}%
\providecommand \natexlab [1]{#1}%
\providecommand \enquote  [1]{``#1''}%
\providecommand \bibnamefont  [1]{#1}%
\providecommand \bibfnamefont [1]{#1}%
\providecommand \citenamefont [1]{#1}%
\providecommand \href@noop [0]{\@secondoftwo}%
\providecommand \href [0]{\begingroup \@sanitize@url \@href}%
\providecommand \@href[1]{\@@startlink{#1}\@@href}%
\providecommand \@@href[1]{\endgroup#1\@@endlink}%
\providecommand \@sanitize@url [0]{\catcode `\\12\catcode `\$12\catcode
  `\&12\catcode `\#12\catcode `\^12\catcode `\_12\catcode `\%12\relax}%
\providecommand \@@startlink[1]{}%
\providecommand \@@endlink[0]{}%
\providecommand \url  [0]{\begingroup\@sanitize@url \@url }%
\providecommand \@url [1]{\endgroup\@href {#1}{\urlprefix }}%
\providecommand \urlprefix  [0]{URL }%
\providecommand \Eprint [0]{\href }%
\providecommand \doibase [0]{https://doi.org/}%
\providecommand \selectlanguage [0]{\@gobble}%
\providecommand \bibinfo  [0]{\@secondoftwo}%
\providecommand \bibfield  [0]{\@secondoftwo}%
\providecommand \translation [1]{[#1]}%
\providecommand \BibitemOpen [0]{}%
\providecommand \bibitemStop [0]{}%
\providecommand \bibitemNoStop [0]{.\EOS\space}%
\providecommand \EOS [0]{\spacefactor3000\relax}%
\providecommand \BibitemShut  [1]{\csname bibitem#1\endcsname}%
\let\auto@bib@innerbib\@empty
\bibitem [{\citenamefont {Cover}\ and\ \citenamefont
  {Thomas}(2006)}]{CoverThomas}%
  \BibitemOpen
  \bibfield  {author} {\bibinfo {author} {\bibfnamefont {T.~M.}\ \bibnamefont
  {Cover}}\ and\ \bibinfo {author} {\bibfnamefont {J.~A.}\ \bibnamefont
  {Thomas}},\ }\href@noop {} {\emph {\bibinfo {title} {Elements of Information
  Theory (Wiley Series in Telecommunications and Signal Processing)}}}\
  (\bibinfo  {publisher} {Wiley-Interscience},\ \bibinfo {address} {USA},\
  \bibinfo {year} {2006})\BibitemShut {NoStop}%
\bibitem [{\citenamefont {Nielsen}\ and\ \citenamefont
  {Chuang}(2000)}]{Nielsen:Book}%
  \BibitemOpen
  \bibfield  {author} {\bibinfo {author} {\bibfnamefont {M.}~\bibnamefont
  {Nielsen}}\ and\ \bibinfo {author} {\bibfnamefont {I.}~\bibnamefont
  {Chuang}},\ }\href {https://books.google.pl/books?id=65FqEKQOfP8C} {\emph
  {\bibinfo {title} {Quantum Computation and Quantum Information}}},\ Cambridge
  Series on Information and the Natural Sciences\ (\bibinfo  {publisher}
  {Cambridge University Press},\ \bibinfo {year} {2000})\BibitemShut {NoStop}%
\bibitem [{\citenamefont {Salles}\ \emph {et~al.}(2008)\citenamefont {Salles},
  \citenamefont {de~Melo}, \citenamefont {Almeida}, \citenamefont {Hor-Meyll},
  \citenamefont {Walborn}, \citenamefont {Souto~Ribeiro},\ and\ \citenamefont
  {Davidovich}}]{PRA.78.Davidovich}%
  \BibitemOpen
  \bibfield  {author} {\bibinfo {author} {\bibfnamefont {A.}~\bibnamefont
  {Salles}}, \bibinfo {author} {\bibfnamefont {F.}~\bibnamefont {de~Melo}},
  \bibinfo {author} {\bibfnamefont {M.~P.}\ \bibnamefont {Almeida}}, \bibinfo
  {author} {\bibfnamefont {M.}~\bibnamefont {Hor-Meyll}}, \bibinfo {author}
  {\bibfnamefont {S.~P.}\ \bibnamefont {Walborn}}, \bibinfo {author}
  {\bibfnamefont {P.~H.}\ \bibnamefont {Souto~Ribeiro}},\ and\ \bibinfo
  {author} {\bibfnamefont {L.}~\bibnamefont {Davidovich}},\ }\href
  {https://doi.org/10.1103/PhysRevA.78.022322} {\bibfield  {journal} {\bibinfo
  {journal} {Phys. Rev. A}\ }\textbf {\bibinfo {volume} {78}},\ \bibinfo
  {pages} {022322} (\bibinfo {year} {2008})}\BibitemShut {NoStop}%
\bibitem [{\citenamefont {Souza}\ \emph {et~al.}(2007)\citenamefont {Souza},
  \citenamefont {Huguenin}, \citenamefont {Milman},\ and\ \citenamefont
  {Khoury}}]{PRL.99.Topo}%
  \BibitemOpen
  \bibfield  {author} {\bibinfo {author} {\bibfnamefont {C.~E.~R.}\
  \bibnamefont {Souza}}, \bibinfo {author} {\bibfnamefont {J.~A.~O.}\
  \bibnamefont {Huguenin}}, \bibinfo {author} {\bibfnamefont {P.}~\bibnamefont
  {Milman}},\ and\ \bibinfo {author} {\bibfnamefont {A.~Z.}\ \bibnamefont
  {Khoury}},\ }\href {https://doi.org/10.1103/PhysRevLett.99.160401} {\bibfield
   {journal} {\bibinfo  {journal} {Phys. Rev. Lett.}\ }\textbf {\bibinfo
  {volume} {99}},\ \bibinfo {pages} {160401} (\bibinfo {year}
  {2007})}\BibitemShut {NoStop}%
\bibitem [{\citenamefont {Qian}\ \emph {et~al.}(2017)\citenamefont {Qian},
  \citenamefont {Vamivakas},\ and\ \citenamefont {Eberly}}]{Eberly}%
  \BibitemOpen
  \bibfield  {author} {\bibinfo {author} {\bibfnamefont {X.-F.}\ \bibnamefont
  {Qian}}, \bibinfo {author} {\bibfnamefont {A.~N.}\ \bibnamefont
  {Vamivakas}},\ and\ \bibinfo {author} {\bibfnamefont {J.~H.}\ \bibnamefont
  {Eberly}},\ }\href {https://doi.org/10.1364/OPN.28.10.000034} {\bibfield
  {journal} {\bibinfo  {journal} {Opt. Photon. News}\ }\textbf {\bibinfo
  {volume} {28}},\ \bibinfo {pages} {34} (\bibinfo {year} {2017})}\BibitemShut
  {NoStop}%
\bibitem [{\citenamefont {Borges}\ \emph {et~al.}(2010)\citenamefont {Borges},
  \citenamefont {Hor-Meyll}, \citenamefont {Huguenin},\ and\ \citenamefont
  {Khoury}}]{PRA.82.Borges}%
  \BibitemOpen
  \bibfield  {author} {\bibinfo {author} {\bibfnamefont {C.~V.~S.}\
  \bibnamefont {Borges}}, \bibinfo {author} {\bibfnamefont {M.}~\bibnamefont
  {Hor-Meyll}}, \bibinfo {author} {\bibfnamefont {J.~A.~O.}\ \bibnamefont
  {Huguenin}},\ and\ \bibinfo {author} {\bibfnamefont {A.~Z.}\ \bibnamefont
  {Khoury}},\ }\href {https://doi.org/10.1103/PhysRevA.82.033833} {\bibfield
  {journal} {\bibinfo  {journal} {Phys. Rev. A}\ }\textbf {\bibinfo {volume}
  {82}},\ \bibinfo {pages} {033833} (\bibinfo {year} {2010})}\BibitemShut
  {NoStop}%
\bibitem [{\citenamefont {Kagalwala}\ \emph {et~al.}(2013)\citenamefont
  {Kagalwala}, \citenamefont {Di~Giuseppe}, \citenamefont {Abouraddy},\ and\
  \citenamefont {Saleh}}]{NaturePhot.Kagalwala-2012}%
  \BibitemOpen
  \bibfield  {author} {\bibinfo {author} {\bibfnamefont {K.~H.}\ \bibnamefont
  {Kagalwala}}, \bibinfo {author} {\bibfnamefont {G.}~\bibnamefont
  {Di~Giuseppe}}, \bibinfo {author} {\bibfnamefont {A.~F.}\ \bibnamefont
  {Abouraddy}},\ and\ \bibinfo {author} {\bibfnamefont {B.~E.~A.}\ \bibnamefont
  {Saleh}},\ }\href {https://doi.org/10.1038/nphoton.2012.312} {\bibfield
  {journal} {\bibinfo  {journal} {Nature Photonics}\ }\textbf {\bibinfo
  {volume} {7}},\ \bibinfo {pages} {72} (\bibinfo {year} {2013})}\BibitemShut
  {NoStop}%
\bibitem [{\citenamefont {Balthazar}\ \emph {et~al.}(2016)\citenamefont
  {Balthazar}, \citenamefont {Souza}, \citenamefont {Caetano}, \citenamefont
  {{a}o}, \citenamefont {Huguenin},\ and\ \citenamefont
  {Khoury}}]{Opt.Lett.Balthazar-2016}%
  \BibitemOpen
  \bibfield  {author} {\bibinfo {author} {\bibfnamefont {W.~F.}\ \bibnamefont
  {Balthazar}}, \bibinfo {author} {\bibfnamefont {C.~E.~R.}\ \bibnamefont
  {Souza}}, \bibinfo {author} {\bibfnamefont {D.~P.}\ \bibnamefont {Caetano}},
  \bibinfo {author} {\bibfnamefont {E.~F.~G.}\ \bibnamefont {{a}o}}, \bibinfo
  {author} {\bibfnamefont {J.~A.~O.}\ \bibnamefont {Huguenin}},\ and\ \bibinfo
  {author} {\bibfnamefont {A.~Z.}\ \bibnamefont {Khoury}},\ }\href
  {https://doi.org/10.1364/OL.41.005797} {\bibfield  {journal} {\bibinfo
  {journal} {Opt. Lett.}\ }\textbf {\bibinfo {volume} {41}},\ \bibinfo {pages}
  {5797} (\bibinfo {year} {2016})}\BibitemShut {NoStop}%
\bibitem [{\citenamefont {Souza}\ \emph {et~al.}(2008)\citenamefont {Souza},
  \citenamefont {Borges}, \citenamefont {Khoury}, \citenamefont {Huguenin},
  \citenamefont {Aolita},\ and\ \citenamefont {Walborn}}]{PRA.77.Cadu-Cripto}%
  \BibitemOpen
  \bibfield  {author} {\bibinfo {author} {\bibfnamefont {C.~E.~R.}\
  \bibnamefont {Souza}}, \bibinfo {author} {\bibfnamefont {C.~V.~S.}\
  \bibnamefont {Borges}}, \bibinfo {author} {\bibfnamefont {A.~Z.}\
  \bibnamefont {Khoury}}, \bibinfo {author} {\bibfnamefont {J.~A.~O.}\
  \bibnamefont {Huguenin}}, \bibinfo {author} {\bibfnamefont {L.}~\bibnamefont
  {Aolita}},\ and\ \bibinfo {author} {\bibfnamefont {S.~P.}\ \bibnamefont
  {Walborn}},\ }\href {https://doi.org/10.1103/PhysRevA.77.032345} {\bibfield
  {journal} {\bibinfo  {journal} {Phys. Rev. A}\ }\textbf {\bibinfo {volume}
  {77}},\ \bibinfo {pages} {032345} (\bibinfo {year} {2008})}\BibitemShut
  {NoStop}%
\bibitem [{\citenamefont {Khoury}\ and\ \citenamefont
  {Milman}(2011)}]{PRA.83.Zela-Teleport}%
  \BibitemOpen
  \bibfield  {author} {\bibinfo {author} {\bibfnamefont {A.~Z.}\ \bibnamefont
  {Khoury}}\ and\ \bibinfo {author} {\bibfnamefont {P.}~\bibnamefont
  {Milman}},\ }\href {https://doi.org/10.1103/PhysRevA.83.060301} {\bibfield
  {journal} {\bibinfo  {journal} {Phys. Rev. A}\ }\textbf {\bibinfo {volume}
  {83}},\ \bibinfo {pages} {060301} (\bibinfo {year} {2011})}\BibitemShut
  {NoStop}%
\bibitem [{\citenamefont {Souza}\ and\ \citenamefont
  {Khoury}(2010)}]{Opt.Exp.Souza-Cnot-2010}%
  \BibitemOpen
  \bibfield  {author} {\bibinfo {author} {\bibfnamefont {C.~E.~R.}\
  \bibnamefont {Souza}}\ and\ \bibinfo {author} {\bibfnamefont {A.~Z.}\
  \bibnamefont {Khoury}},\ }\href {https://doi.org/10.1364/OE.18.009207}
  {\bibfield  {journal} {\bibinfo  {journal} {Opt. Express}\ }\textbf {\bibinfo
  {volume} {18}},\ \bibinfo {pages} {9207} (\bibinfo {year}
  {2010})}\BibitemShut {NoStop}%
\bibitem [{\citenamefont {Balthazar}\ and\ \citenamefont
  {Huguenin}(2016)}]{JOPS-B.Cod.Op.Balthazar-2016}%
  \BibitemOpen
  \bibfield  {author} {\bibinfo {author} {\bibfnamefont {W.~F.}\ \bibnamefont
  {Balthazar}}\ and\ \bibinfo {author} {\bibfnamefont {J.~A.~O.}\ \bibnamefont
  {Huguenin}},\ }\href {https://doi.org/10.1364/JOSAB.33.001649} {\bibfield
  {journal} {\bibinfo  {journal} {J. Opt. Soc. Am. B}\ }\textbf {\bibinfo
  {volume} {33}},\ \bibinfo {pages} {1649} (\bibinfo {year}
  {2016})}\BibitemShut {NoStop}%
\bibitem [{\citenamefont {Passos}\ \emph {et~al.}(2018)\citenamefont {Passos},
  \citenamefont {Balthazar}, \citenamefont {Khoury}, \citenamefont {Hor-Meyll},
  \citenamefont {Davidovich},\ and\ \citenamefont
  {Huguenin}}]{PRA.97.Enviro-Passos}%
  \BibitemOpen
  \bibfield  {author} {\bibinfo {author} {\bibfnamefont {M.~H.~M.}\
  \bibnamefont {Passos}}, \bibinfo {author} {\bibfnamefont {W.~F.}\
  \bibnamefont {Balthazar}}, \bibinfo {author} {\bibfnamefont {A.~Z.}\
  \bibnamefont {Khoury}}, \bibinfo {author} {\bibfnamefont {M.}~\bibnamefont
  {Hor-Meyll}}, \bibinfo {author} {\bibfnamefont {L.}~\bibnamefont
  {Davidovich}},\ and\ \bibinfo {author} {\bibfnamefont {J.~A.~O.}\
  \bibnamefont {Huguenin}},\ }\href
  {https://doi.org/10.1103/PhysRevA.97.022321} {\bibfield  {journal} {\bibinfo
  {journal} {Phys. Rev. A}\ }\textbf {\bibinfo {volume} {97}},\ \bibinfo
  {pages} {022321} (\bibinfo {year} {2018})}\BibitemShut {NoStop}%
\bibitem [{\citenamefont {Preskill}(2015)}]{Preskill}%
  \BibitemOpen
  \bibfield  {author} {\bibinfo {author} {\bibfnamefont {J.}~\bibnamefont
  {Preskill}},\ }\href {https://books.google.pl/books?id=MIv8rQEACAAJ} {\emph
  {\bibinfo {title} {Lecture Notes for Physics 229:Quantum Information and
  Computation}}}\ (\bibinfo  {publisher} {CreateSpace Independent Publishing
  Platform},\ \bibinfo {year} {2015})\BibitemShut {NoStop}%
\bibitem [{\citenamefont {Kitaev}(1997)}]{Kitaev_1997}%
  \BibitemOpen
  \bibfield  {author} {\bibinfo {author} {\bibfnamefont {A.~Y.}\ \bibnamefont
  {Kitaev}},\ }\href {https://doi.org/10.1070/rm1997v052n06abeh002155}
  {\bibfield  {journal} {\bibinfo  {journal} {Russian Mathematical Surveys}\
  }\textbf {\bibinfo {volume} {52}},\ \bibinfo {pages} {1191} (\bibinfo {year}
  {1997})}\BibitemShut {NoStop}%
\bibitem [{\citenamefont {Dawson}\ and\ \citenamefont {Nielsen}(2006)}]{DN}%
  \BibitemOpen
  \bibfield  {author} {\bibinfo {author} {\bibfnamefont {C.~M.}\ \bibnamefont
  {Dawson}}\ and\ \bibinfo {author} {\bibfnamefont {M.~A.}\ \bibnamefont
  {Nielsen}},\ }\href@noop {} {\bibfield  {journal} {\bibinfo  {journal}
  {Quantum Info. Comput.}\ }\textbf {\bibinfo {volume} {6}},\ \bibinfo {pages}
  {81–95} (\bibinfo {year} {2006})}\BibitemShut {NoStop}%
\bibitem [{\citenamefont {Wang}\ \emph {et~al.}(2013)\citenamefont {Wang},
  \citenamefont {Berry}, \citenamefont {de~Oliveira},\ and\ \citenamefont
  {Sanders}}]{PhysRevLett.111.130504}%
  \BibitemOpen
  \bibfield  {author} {\bibinfo {author} {\bibfnamefont {D.-S.}\ \bibnamefont
  {Wang}}, \bibinfo {author} {\bibfnamefont {D.~W.}\ \bibnamefont {Berry}},
  \bibinfo {author} {\bibfnamefont {M.~C.}\ \bibnamefont {de~Oliveira}},\ and\
  \bibinfo {author} {\bibfnamefont {B.~C.}\ \bibnamefont {Sanders}},\ }\href
  {https://doi.org/10.1103/PhysRevLett.111.130504} {\bibfield  {journal}
  {\bibinfo  {journal} {Phys. Rev. Lett.}\ }\textbf {\bibinfo {volume} {111}},\
  \bibinfo {pages} {130504} (\bibinfo {year} {2013})}\BibitemShut {NoStop}%
\bibitem [{\citenamefont {Wang}\ and\ \citenamefont
  {Sanders}(2015)}]{Wang_2015}%
  \BibitemOpen
  \bibfield  {author} {\bibinfo {author} {\bibfnamefont {D.-S.}\ \bibnamefont
  {Wang}}\ and\ \bibinfo {author} {\bibfnamefont {B.~C.}\ \bibnamefont
  {Sanders}},\ }\href {https://doi.org/10.1088/1367-2630/17/4/043004}
  {\bibfield  {journal} {\bibinfo  {journal} {New Journal of Physics}\ }\textbf
  {\bibinfo {volume} {17}},\ \bibinfo {pages} {043004} (\bibinfo {year}
  {2015})}\BibitemShut {NoStop}%
\bibitem [{\citenamefont {Hu}\ \emph {et~al.}(2018)\citenamefont {Hu},
  \citenamefont {Mu}, \citenamefont {Cai}, \citenamefont {Ma}, \citenamefont
  {Xu}, \citenamefont {Wang}, \citenamefont {Song}, \citenamefont {Zou},\ and\
  \citenamefont {Sun}}]{Hu-Exp_SK:18}%
  \BibitemOpen
  \bibfield  {author} {\bibinfo {author} {\bibfnamefont {L.}~\bibnamefont
  {Hu}}, \bibinfo {author} {\bibfnamefont {X.}~\bibnamefont {Mu}}, \bibinfo
  {author} {\bibfnamefont {W.}~\bibnamefont {Cai}}, \bibinfo {author}
  {\bibfnamefont {Y.}~\bibnamefont {Ma}}, \bibinfo {author} {\bibfnamefont
  {Y.}~\bibnamefont {Xu}}, \bibinfo {author} {\bibfnamefont {H.}~\bibnamefont
  {Wang}}, \bibinfo {author} {\bibfnamefont {Y.}~\bibnamefont {Song}}, \bibinfo
  {author} {\bibfnamefont {C.-L.}\ \bibnamefont {Zou}},\ and\ \bibinfo {author}
  {\bibfnamefont {L.}~\bibnamefont {Sun}},\ }\href
  {https://doi.org/https://doi.org/10.1016/j.scib.2018.11.010} {\bibfield
  {journal} {\bibinfo  {journal} {Science Bulletin}\ }\textbf {\bibinfo
  {volume} {63}},\ \bibinfo {pages} {1551 } (\bibinfo {year}
  {2018})}\BibitemShut {NoStop}%
\bibitem [{\citenamefont {Xin}\ \emph {et~al.}(2017)\citenamefont {Xin},
  \citenamefont {Wei}, \citenamefont {Pedernales}, \citenamefont {Solano},\
  and\ \citenamefont {Long}}]{PhysRevA.96.062303}%
  \BibitemOpen
  \bibfield  {author} {\bibinfo {author} {\bibfnamefont {T.}~\bibnamefont
  {Xin}}, \bibinfo {author} {\bibfnamefont {S.-J.}\ \bibnamefont {Wei}},
  \bibinfo {author} {\bibfnamefont {J.~S.}\ \bibnamefont {Pedernales}},
  \bibinfo {author} {\bibfnamefont {E.}~\bibnamefont {Solano}},\ and\ \bibinfo
  {author} {\bibfnamefont {G.-L.}\ \bibnamefont {Long}},\ }\href
  {https://doi.org/10.1103/PhysRevA.96.062303} {\bibfield  {journal} {\bibinfo
  {journal} {Phys. Rev. A}\ }\textbf {\bibinfo {volume} {96}},\ \bibinfo
  {pages} {062303} (\bibinfo {year} {2017})}\BibitemShut {NoStop}%
\bibitem [{\citenamefont {McCutcheon}\ \emph {et~al.}(2018)\citenamefont
  {McCutcheon}, \citenamefont {McMillan}, \citenamefont {Rarity},\ and\
  \citenamefont {Tame}}]{McCutcheon-EXP_SK:18}%
  \BibitemOpen
  \bibfield  {author} {\bibinfo {author} {\bibfnamefont {W.}~\bibnamefont
  {McCutcheon}}, \bibinfo {author} {\bibfnamefont {A.}~\bibnamefont
  {McMillan}}, \bibinfo {author} {\bibfnamefont {J.~G.}\ \bibnamefont
  {Rarity}},\ and\ \bibinfo {author} {\bibfnamefont {M.~S.}\ \bibnamefont
  {Tame}},\ }\href {https://doi.org/10.1088/1367-2630/aa9b5c} {\bibfield
  {journal} {\bibinfo  {journal} {New Journal of Physics}\ }\textbf {\bibinfo
  {volume} {20}},\ \bibinfo {pages} {033019} (\bibinfo {year}
  {2018})}\BibitemShut {NoStop}%
\bibitem [{\citenamefont {Lu}\ \emph {et~al.}(2017)\citenamefont {Lu},
  \citenamefont {Liu}, \citenamefont {Wang}, \citenamefont {Chen},
  \citenamefont {Li}, \citenamefont {Yao}, \citenamefont {Li}, \citenamefont
  {Liu}, \citenamefont {Peng}, \citenamefont {Sanders}, \citenamefont {Chen},\
  and\ \citenamefont {Pan}}]{SK-Lu_Wei:17}%
  \BibitemOpen
  \bibfield  {author} {\bibinfo {author} {\bibfnamefont {H.}~\bibnamefont
  {Lu}}, \bibinfo {author} {\bibfnamefont {C.}~\bibnamefont {Liu}}, \bibinfo
  {author} {\bibfnamefont {D.-S.}\ \bibnamefont {Wang}}, \bibinfo {author}
  {\bibfnamefont {L.-K.}\ \bibnamefont {Chen}}, \bibinfo {author}
  {\bibfnamefont {Z.-D.}\ \bibnamefont {Li}}, \bibinfo {author} {\bibfnamefont
  {X.-C.}\ \bibnamefont {Yao}}, \bibinfo {author} {\bibfnamefont
  {L.}~\bibnamefont {Li}}, \bibinfo {author} {\bibfnamefont {N.-L.}\
  \bibnamefont {Liu}}, \bibinfo {author} {\bibfnamefont {C.-Z.}\ \bibnamefont
  {Peng}}, \bibinfo {author} {\bibfnamefont {B.~C.}\ \bibnamefont {Sanders}},
  \bibinfo {author} {\bibfnamefont {Y.-A.}\ \bibnamefont {Chen}},\ and\
  \bibinfo {author} {\bibfnamefont {J.-W.}\ \bibnamefont {Pan}},\ }\href
  {https://doi.org/10.1103/PhysRevA.95.042310} {\bibfield  {journal} {\bibinfo
  {journal} {Phys. Rev. A}\ }\textbf {\bibinfo {volume} {95}},\ \bibinfo
  {pages} {042310} (\bibinfo {year} {2017})}\BibitemShut {NoStop}%
\bibitem [{\citenamefont {Marrucci}\ \emph {et~al.}(2012)\citenamefont
  {Marrucci}, \citenamefont {Karimi}, \citenamefont {Slussarenko},
  \citenamefont {Piccirillo}, \citenamefont {Santamato}, \citenamefont
  {Nagali},\ and\ \citenamefont {Sciarrino}}]{Marrucci12}%
  \BibitemOpen
  \bibfield  {author} {\bibinfo {author} {\bibfnamefont {L.}~\bibnamefont
  {Marrucci}}, \bibinfo {author} {\bibfnamefont {E.}~\bibnamefont {Karimi}},
  \bibinfo {author} {\bibfnamefont {S.}~\bibnamefont {Slussarenko}}, \bibinfo
  {author} {\bibfnamefont {B.}~\bibnamefont {Piccirillo}}, \bibinfo {author}
  {\bibfnamefont {E.}~\bibnamefont {Santamato}}, \bibinfo {author}
  {\bibfnamefont {E.}~\bibnamefont {Nagali}},\ and\ \bibinfo {author}
  {\bibfnamefont {F.}~\bibnamefont {Sciarrino}},\ }\href
  {https://doi.org/10.1080/15421406.2012.686710} {\bibfield  {journal}
  {\bibinfo  {journal} {Molecular Crystals and Liquid Crystals}\ }\textbf
  {\bibinfo {volume} {561}},\ \bibinfo {pages} {48} (\bibinfo {year}
  {2012})}\BibitemShut {NoStop}%
\bibitem [{\citenamefont {Cardano}\ \emph {et~al.}(2012)\citenamefont
  {Cardano}, \citenamefont {Karimi}, \citenamefont {Slussarenko}, \citenamefont
  {Marrucci}, \citenamefont {de~Lisio},\ and\ \citenamefont
  {Santamato}}]{Cardano12}%
  \BibitemOpen
  \bibfield  {author} {\bibinfo {author} {\bibfnamefont {F.}~\bibnamefont
  {Cardano}}, \bibinfo {author} {\bibfnamefont {E.}~\bibnamefont {Karimi}},
  \bibinfo {author} {\bibfnamefont {S.}~\bibnamefont {Slussarenko}}, \bibinfo
  {author} {\bibfnamefont {L.}~\bibnamefont {Marrucci}}, \bibinfo {author}
  {\bibfnamefont {C.}~\bibnamefont {de~Lisio}},\ and\ \bibinfo {author}
  {\bibfnamefont {E.}~\bibnamefont {Santamato}},\ }\href
  {https://doi.org/10.1364/AO.51.0000C1} {\bibfield  {journal} {\bibinfo
  {journal} {Appl. Opt.}\ }\textbf {\bibinfo {volume} {51}},\ \bibinfo {pages}
  {C1} (\bibinfo {year} {2012})}\BibitemShut {NoStop}%
\bibitem [{\citenamefont {Filippo~Cardano}(2001)}]{Cardano01}%
  \BibitemOpen
  \bibfield  {author} {\bibinfo {author} {\bibfnamefont {L.~M.}\ \bibnamefont
  {Filippo~Cardano}},\ }\href {https://doi.org/10.1038/nphoton.2015.232}
  {\bibfield  {journal} {\bibinfo  {journal} {Nature Photonics}\ }\textbf
  {\bibinfo {volume} {9}},\ \bibinfo {pages} {C1} (\bibinfo {year}
  {2001})}\BibitemShut {NoStop}%
\bibitem [{\citenamefont {Karimi}\ \emph {et~al.}(2012)\citenamefont {Karimi},
  \citenamefont {Marrucci}, \citenamefont {Grillo},\ and\ \citenamefont
  {Santamato}}]{Karimi12}%
  \BibitemOpen
  \bibfield  {author} {\bibinfo {author} {\bibfnamefont {E.}~\bibnamefont
  {Karimi}}, \bibinfo {author} {\bibfnamefont {L.}~\bibnamefont {Marrucci}},
  \bibinfo {author} {\bibfnamefont {V.}~\bibnamefont {Grillo}},\ and\ \bibinfo
  {author} {\bibfnamefont {E.}~\bibnamefont {Santamato}},\ }\href
  {https://doi.org/10.1103/PhysRevLett.108.044801} {\bibfield  {journal}
  {\bibinfo  {journal} {Phys. Rev. Lett.}\ }\textbf {\bibinfo {volume} {108}},\
  \bibinfo {pages} {044801} (\bibinfo {year} {2012})}\BibitemShut {NoStop}%
\bibitem [{\citenamefont {Nagali}\ \emph {et~al.}(2009)\citenamefont {Nagali},
  \citenamefont {Sciarrino}, \citenamefont {De~Martini}, \citenamefont
  {Marrucci}, \citenamefont {Piccirillo}, \citenamefont {Karimi},\ and\
  \citenamefont {Santamato}}]{Nagali09}%
  \BibitemOpen
  \bibfield  {author} {\bibinfo {author} {\bibfnamefont {E.}~\bibnamefont
  {Nagali}}, \bibinfo {author} {\bibfnamefont {F.}~\bibnamefont {Sciarrino}},
  \bibinfo {author} {\bibfnamefont {F.}~\bibnamefont {De~Martini}}, \bibinfo
  {author} {\bibfnamefont {L.}~\bibnamefont {Marrucci}}, \bibinfo {author}
  {\bibfnamefont {B.}~\bibnamefont {Piccirillo}}, \bibinfo {author}
  {\bibfnamefont {E.}~\bibnamefont {Karimi}},\ and\ \bibinfo {author}
  {\bibfnamefont {E.}~\bibnamefont {Santamato}},\ }\href
  {https://doi.org/10.1103/PhysRevLett.103.013601} {\bibfield  {journal}
  {\bibinfo  {journal} {Phys. Rev. Lett.}\ }\textbf {\bibinfo {volume} {103}},\
  \bibinfo {pages} {013601} (\bibinfo {year} {2009})}\BibitemShut {NoStop}%
\bibitem [{\citenamefont {de~Oliveira}\ \emph {et~al.}(2020)\citenamefont
  {de~Oliveira}, \citenamefont {Nape}, \citenamefont {Pinnell}, \citenamefont
  {TabeBordbar},\ and\ \citenamefont {Forbes}}]{DeOliveira20}%
  \BibitemOpen
  \bibfield  {author} {\bibinfo {author} {\bibfnamefont {M.}~\bibnamefont
  {de~Oliveira}}, \bibinfo {author} {\bibfnamefont {I.}~\bibnamefont {Nape}},
  \bibinfo {author} {\bibfnamefont {J.}~\bibnamefont {Pinnell}}, \bibinfo
  {author} {\bibfnamefont {N.}~\bibnamefont {TabeBordbar}},\ and\ \bibinfo
  {author} {\bibfnamefont {A.}~\bibnamefont {Forbes}},\ }\href
  {https://doi.org/10.1103/PhysRevA.101.042303} {\bibfield  {journal} {\bibinfo
   {journal} {Phys. Rev. A}\ }\textbf {\bibinfo {volume} {101}},\ \bibinfo
  {pages} {042303} (\bibinfo {year} {2020})}\BibitemShut {NoStop}%
\bibitem [{\citenamefont {Goyal}\ \emph {et~al.}(2013)\citenamefont {Goyal},
  \citenamefont {Roux}, \citenamefont {Forbes},\ and\ \citenamefont
  {Konrad}}]{Goyal13}%
  \BibitemOpen
  \bibfield  {author} {\bibinfo {author} {\bibfnamefont {S.~K.}\ \bibnamefont
  {Goyal}}, \bibinfo {author} {\bibfnamefont {F.~S.}\ \bibnamefont {Roux}},
  \bibinfo {author} {\bibfnamefont {A.}~\bibnamefont {Forbes}},\ and\ \bibinfo
  {author} {\bibfnamefont {T.}~\bibnamefont {Konrad}},\ }\href
  {https://doi.org/10.1103/PhysRevLett.110.263602} {\bibfield  {journal}
  {\bibinfo  {journal} {Phys. Rev. Lett.}\ }\textbf {\bibinfo {volume} {110}},\
  \bibinfo {pages} {263602} (\bibinfo {year} {2013})}\BibitemShut {NoStop}%
\bibitem [{\citenamefont {Hamadou~Ibrahim}\ \emph {et~al.}(2013)\citenamefont
  {Hamadou~Ibrahim}, \citenamefont {Roux}, \citenamefont {McLaren},
  \citenamefont {Konrad},\ and\ \citenamefont {Forbes}}]{Hamadou13}%
  \BibitemOpen
  \bibfield  {author} {\bibinfo {author} {\bibfnamefont {A.}~\bibnamefont
  {Hamadou~Ibrahim}}, \bibinfo {author} {\bibfnamefont {F.~S.}\ \bibnamefont
  {Roux}}, \bibinfo {author} {\bibfnamefont {M.}~\bibnamefont {McLaren}},
  \bibinfo {author} {\bibfnamefont {T.}~\bibnamefont {Konrad}},\ and\ \bibinfo
  {author} {\bibfnamefont {A.}~\bibnamefont {Forbes}},\ }\href
  {https://doi.org/10.1103/PhysRevA.88.012312} {\bibfield  {journal} {\bibinfo
  {journal} {Phys. Rev. A}\ }\textbf {\bibinfo {volume} {88}},\ \bibinfo
  {pages} {012312} (\bibinfo {year} {2013})}\BibitemShut {NoStop}%
\bibitem [{\citenamefont {Konrad}\ and\ \citenamefont
  {Forbes}(2019)}]{Konrad19}%
  \BibitemOpen
  \bibfield  {author} {\bibinfo {author} {\bibfnamefont {T.}~\bibnamefont
  {Konrad}}\ and\ \bibinfo {author} {\bibfnamefont {A.}~\bibnamefont
  {Forbes}},\ }\href {https://doi.org/10.1080/00107514.2019.1580433} {\bibfield
   {journal} {\bibinfo  {journal} {Contemporary Physics}\ }\textbf {\bibinfo
  {volume} {60}},\ \bibinfo {pages} {1} (\bibinfo {year} {2019})},\ \Eprint
  {https://arxiv.org/abs/https://doi.org/10.1080/00107514.2019.1580433}
  {https://doi.org/10.1080/00107514.2019.1580433} \BibitemShut {NoStop}%
\bibitem [{\citenamefont {McLaren}\ \emph {et~al.}(2015)\citenamefont
  {McLaren}, \citenamefont {Konrad},\ and\ \citenamefont {Forbes}}]{McLaren15}%
  \BibitemOpen
  \bibfield  {author} {\bibinfo {author} {\bibfnamefont {M.}~\bibnamefont
  {McLaren}}, \bibinfo {author} {\bibfnamefont {T.}~\bibnamefont {Konrad}},\
  and\ \bibinfo {author} {\bibfnamefont {A.}~\bibnamefont {Forbes}},\ }\href
  {https://doi.org/10.1103/PhysRevA.92.023833} {\bibfield  {journal} {\bibinfo
  {journal} {Phys. Rev. A}\ }\textbf {\bibinfo {volume} {92}},\ \bibinfo
  {pages} {023833} (\bibinfo {year} {2015})}\BibitemShut {NoStop}%
\bibitem [{\citenamefont {Gailele}\ \emph {et~al.}(2018)\citenamefont
  {Gailele}, \citenamefont {Dudley},\ and\ \citenamefont {Forbes}}]{Gailele18}%
  \BibitemOpen
  \bibfield  {author} {\bibinfo {author} {\bibfnamefont {L.}~\bibnamefont
  {Gailele}}, \bibinfo {author} {\bibfnamefont {A.}~\bibnamefont {Dudley}},\
  and\ \bibinfo {author} {\bibfnamefont {A.}~\bibnamefont {Forbes}},\ }in\
  \href {https://doi.org/10.1117/12.2505725} {\emph {\bibinfo {booktitle}
  {Laser Beam Shaping XVIII}}},\ Vol.\ \bibinfo {volume} {10744},\ \bibinfo
  {editor} {edited by\ \bibinfo {editor} {\bibfnamefont {A.}~\bibnamefont
  {Dudley}}\ and\ \bibinfo {editor} {\bibfnamefont {A.~V.}\ \bibnamefont
  {Laskin}}},\ \bibinfo {organization} {International Society for Optics and
  Photonics}\ (\bibinfo  {publisher} {SPIE},\ \bibinfo {year} {2018})\ pp.\
  \bibinfo {pages} {210 -- 215}\BibitemShut {NoStop}%
\bibitem [{\citenamefont {Johnson}\ \emph {et~al.}(2019)\citenamefont
  {Johnson}, \citenamefont {Ma}, \citenamefont {Padgett},\ and\ \citenamefont
  {Ramachandran}}]{Johnson19}%
  \BibitemOpen
  \bibfield  {author} {\bibinfo {author} {\bibfnamefont {S.~D.}\ \bibnamefont
  {Johnson}}, \bibinfo {author} {\bibfnamefont {Z.}~\bibnamefont {Ma}},
  \bibinfo {author} {\bibfnamefont {M.~J.}\ \bibnamefont {Padgett}},\ and\
  \bibinfo {author} {\bibfnamefont {S.}~\bibnamefont {Ramachandran}},\ }\href
  {https://doi.org/10.1364/OSAC.2.002975} {\bibfield  {journal} {\bibinfo
  {journal} {OSA Continuum}\ }\textbf {\bibinfo {volume} {2}},\ \bibinfo
  {pages} {2975} (\bibinfo {year} {2019})}\BibitemShut {NoStop}%
\bibitem [{\citenamefont {Mirhosseini}\ \emph {et~al.}(2015)\citenamefont
  {Mirhosseini}, \citenamefont {Maga{\~{n}}a-Loaiza}, \citenamefont
  {O'Sullivan}, \citenamefont {Rodenburg}, \citenamefont {Malik}, \citenamefont
  {Lavery}, \citenamefont {Padgett}, \citenamefont {Gauthier},\ and\
  \citenamefont {Boyd}}]{Mirhosseini15}%
  \BibitemOpen
  \bibfield  {author} {\bibinfo {author} {\bibfnamefont {M.}~\bibnamefont
  {Mirhosseini}}, \bibinfo {author} {\bibfnamefont {O.~S.}\ \bibnamefont
  {Maga{\~{n}}a-Loaiza}}, \bibinfo {author} {\bibfnamefont {M.~N.}\
  \bibnamefont {O'Sullivan}}, \bibinfo {author} {\bibfnamefont
  {B.}~\bibnamefont {Rodenburg}}, \bibinfo {author} {\bibfnamefont
  {M.}~\bibnamefont {Malik}}, \bibinfo {author} {\bibfnamefont {M.~P.~J.}\
  \bibnamefont {Lavery}}, \bibinfo {author} {\bibfnamefont {M.~J.}\
  \bibnamefont {Padgett}}, \bibinfo {author} {\bibfnamefont {D.~J.}\
  \bibnamefont {Gauthier}},\ and\ \bibinfo {author} {\bibfnamefont {R.~W.}\
  \bibnamefont {Boyd}},\ }\href {https://doi.org/10.1088/1367-2630/17/3/033033}
  {\bibfield  {journal} {\bibinfo  {journal} {New Journal of Physics}\ }\textbf
  {\bibinfo {volume} {17}},\ \bibinfo {pages} {033033} (\bibinfo {year}
  {2015})}\BibitemShut {NoStop}%
\bibitem [{\citenamefont {Mirhosseini}\ \emph {et~al.}(2016)\citenamefont
  {Mirhosseini}, \citenamefont {{n}a Loaiza}, \citenamefont {O'Sullivan},
  \citenamefont {Rodenburg}, \citenamefont {Shi}, \citenamefont {Malik},
  \citenamefont {Lavery}, \citenamefont {Padgett}, \citenamefont {Gauthier},\
  and\ \citenamefont {Boyd}}]{Mirhosseini16}%
  \BibitemOpen
  \bibfield  {author} {\bibinfo {author} {\bibfnamefont {M.}~\bibnamefont
  {Mirhosseini}}, \bibinfo {author} {\bibfnamefont {O.~S.~M.}\ \bibnamefont
  {{n}a Loaiza}}, \bibinfo {author} {\bibfnamefont {M.~N.}\ \bibnamefont
  {O'Sullivan}}, \bibinfo {author} {\bibfnamefont {B.}~\bibnamefont
  {Rodenburg}}, \bibinfo {author} {\bibfnamefont {Z.}~\bibnamefont {Shi}},
  \bibinfo {author} {\bibfnamefont {M.}~\bibnamefont {Malik}}, \bibinfo
  {author} {\bibfnamefont {M.~P.~J.}\ \bibnamefont {Lavery}}, \bibinfo {author}
  {\bibfnamefont {M.~J.}\ \bibnamefont {Padgett}}, \bibinfo {author}
  {\bibfnamefont {D.~J.}\ \bibnamefont {Gauthier}},\ and\ \bibinfo {author}
  {\bibfnamefont {R.~W.}\ \bibnamefont {Boyd}},\ }in\ \href
  {https://doi.org/10.1364/LS.2016.LTu1E.3} {\emph {\bibinfo {booktitle}
  {Frontiers in Optics 2016}}}\ (\bibinfo  {publisher} {Optical Society of
  America},\ \bibinfo {year} {2016})\ p.\ \bibinfo {pages}
  {LTu1E.3}\BibitemShut {NoStop}%
\bibitem [{\citenamefont {Arlt}\ \emph {et~al.}(1999)\citenamefont {Arlt},
  \citenamefont {Dholakia}, \citenamefont {Allen},\ and\ \citenamefont
  {Padgett}}]{Arlt99}%
  \BibitemOpen
  \bibfield  {author} {\bibinfo {author} {\bibfnamefont {J.}~\bibnamefont
  {Arlt}}, \bibinfo {author} {\bibfnamefont {K.}~\bibnamefont {Dholakia}},
  \bibinfo {author} {\bibfnamefont {L.}~\bibnamefont {Allen}},\ and\ \bibinfo
  {author} {\bibfnamefont {M.~J.}\ \bibnamefont {Padgett}},\ }\href
  {https://doi.org/10.1103/PhysRevA.59.3950} {\bibfield  {journal} {\bibinfo
  {journal} {Phys. Rev. A}\ }\textbf {\bibinfo {volume} {59}},\ \bibinfo
  {pages} {3950} (\bibinfo {year} {1999})}\BibitemShut {NoStop}%
\bibitem [{\citenamefont {Thomaschewski}\ \emph {et~al.}(2019)\citenamefont
  {Thomaschewski}, \citenamefont {Yang}, \citenamefont {Wolff}, \citenamefont
  {Roberts},\ and\ \citenamefont {Bozhevolnyi}}]{Thomaschewski}%
  \BibitemOpen
  \bibfield  {author} {\bibinfo {author} {\bibfnamefont {M.}~\bibnamefont
  {Thomaschewski}}, \bibinfo {author} {\bibfnamefont {Y.}~\bibnamefont {Yang}},
  \bibinfo {author} {\bibfnamefont {C.}~\bibnamefont {Wolff}}, \bibinfo
  {author} {\bibfnamefont {A.~S.}\ \bibnamefont {Roberts}},\ and\ \bibinfo
  {author} {\bibfnamefont {S.~I.}\ \bibnamefont {Bozhevolnyi}},\ }\href
  {https://doi.org/10.1021/acs.nanolett.8b04611} {\bibfield  {journal}
  {\bibinfo  {journal} {Nano Letters}\ }\textbf {\bibinfo {volume} {19}},\
  \bibinfo {pages} {1166} (\bibinfo {year} {2019})}\BibitemShut {NoStop}%
\bibitem [{\citenamefont {Gregg}\ \emph {et~al.}(2019)\citenamefont {Gregg},
  \citenamefont {Kristensen}, \citenamefont {Rubano}, \citenamefont {Golowich},
  \citenamefont {Marrucci},\ and\ \citenamefont {Ramachandran}}]{Gregg}%
  \BibitemOpen
  \bibfield  {author} {\bibinfo {author} {\bibfnamefont {P.}~\bibnamefont
  {Gregg}}, \bibinfo {author} {\bibfnamefont {P.}~\bibnamefont {Kristensen}},
  \bibinfo {author} {\bibfnamefont {A.}~\bibnamefont {Rubano}}, \bibinfo
  {author} {\bibfnamefont {S.}~\bibnamefont {Golowich}}, \bibinfo {author}
  {\bibfnamefont {L.}~\bibnamefont {Marrucci}},\ and\ \bibinfo {author}
  {\bibfnamefont {S.}~\bibnamefont {Ramachandran}},\ }\href
  {https://doi.org/10.1038/s41467-019-12401-4} {\bibfield  {journal} {\bibinfo
  {journal} {Nature Communications}\ }\textbf {\bibinfo {volume} {10}},\
  \bibinfo {pages} {4707} (\bibinfo {year} {2019})}\BibitemShut {NoStop}%
\bibitem [{\citenamefont {Karimi}\ \emph {et~al.}(2010)\citenamefont {Karimi},
  \citenamefont {Leach}, \citenamefont {Slussarenko}, \citenamefont
  {Piccirillo}, \citenamefont {Marrucci}, \citenamefont {Chen}, \citenamefont
  {She}, \citenamefont {Franke-Arnold}, \citenamefont {Padgett},\ and\
  \citenamefont {Santamato}}]{Karimi}%
  \BibitemOpen
  \bibfield  {author} {\bibinfo {author} {\bibfnamefont {E.}~\bibnamefont
  {Karimi}}, \bibinfo {author} {\bibfnamefont {J.}~\bibnamefont {Leach}},
  \bibinfo {author} {\bibfnamefont {S.}~\bibnamefont {Slussarenko}}, \bibinfo
  {author} {\bibfnamefont {B.}~\bibnamefont {Piccirillo}}, \bibinfo {author}
  {\bibfnamefont {L.}~\bibnamefont {Marrucci}}, \bibinfo {author}
  {\bibfnamefont {L.}~\bibnamefont {Chen}}, \bibinfo {author} {\bibfnamefont
  {W.}~\bibnamefont {She}}, \bibinfo {author} {\bibfnamefont {S.}~\bibnamefont
  {Franke-Arnold}}, \bibinfo {author} {\bibfnamefont {M.~J.}\ \bibnamefont
  {Padgett}},\ and\ \bibinfo {author} {\bibfnamefont {E.}~\bibnamefont
  {Santamato}},\ }\href {https://doi.org/10.1103/PhysRevA.82.022115} {\bibfield
   {journal} {\bibinfo  {journal} {Phys. Rev. A}\ }\textbf {\bibinfo {volume}
  {82}},\ \bibinfo {pages} {022115} (\bibinfo {year} {2010})}\BibitemShut
  {NoStop}%
\bibitem [{\citenamefont {Cardano}\ \emph {et~al.}(2013)\citenamefont
  {Cardano}, \citenamefont {Karimi}, \citenamefont {Marrucci}, \citenamefont
  {de~Lisio},\ and\ \citenamefont {Santamato}}]{Cardano}%
  \BibitemOpen
  \bibfield  {author} {\bibinfo {author} {\bibfnamefont {F.}~\bibnamefont
  {Cardano}}, \bibinfo {author} {\bibfnamefont {E.}~\bibnamefont {Karimi}},
  \bibinfo {author} {\bibfnamefont {L.}~\bibnamefont {Marrucci}}, \bibinfo
  {author} {\bibfnamefont {C.}~\bibnamefont {de~Lisio}},\ and\ \bibinfo
  {author} {\bibfnamefont {E.}~\bibnamefont {Santamato}},\ }\href
  {https://doi.org/10.1103/PhysRevA.88.032101} {\bibfield  {journal} {\bibinfo
  {journal} {Phys. Rev. A}\ }\textbf {\bibinfo {volume} {88}},\ \bibinfo
  {pages} {032101} (\bibinfo {year} {2013})}\BibitemShut {NoStop}%
\bibitem [{\citenamefont {D'Ambrosio}\ \emph {et~al.}(2012)\citenamefont
  {D'Ambrosio}, \citenamefont {Nagali}, \citenamefont {Walborn}, \citenamefont
  {Aolita}, \citenamefont {Slussarenko}, \citenamefont {Marrucci},\ and\
  \citenamefont {Sciarrino}}]{marrucci:2012}%
  \BibitemOpen
  \bibfield  {author} {\bibinfo {author} {\bibfnamefont {V.}~\bibnamefont
  {D'Ambrosio}}, \bibinfo {author} {\bibfnamefont {E.}~\bibnamefont {Nagali}},
  \bibinfo {author} {\bibfnamefont {S.~P.}\ \bibnamefont {Walborn}}, \bibinfo
  {author} {\bibfnamefont {L.}~\bibnamefont {Aolita}}, \bibinfo {author}
  {\bibfnamefont {S.}~\bibnamefont {Slussarenko}}, \bibinfo {author}
  {\bibfnamefont {L.}~\bibnamefont {Marrucci}},\ and\ \bibinfo {author}
  {\bibfnamefont {F.}~\bibnamefont {Sciarrino}},\ }\href
  {https://doi.org/10.1038/ncomms1951} {\bibfield  {journal} {\bibinfo
  {journal} {Nature Communications}\ }\textbf {\bibinfo {volume} {3}},\
  \bibinfo {pages} {961} (\bibinfo {year} {2012})}\BibitemShut {NoStop}%
\bibitem [{\citenamefont {Kitaev}\ \emph {et~al.}(2002)\citenamefont {Kitaev},
  \citenamefont {Shen},\ and\ \citenamefont {Vyalyi}}]{Kitaev}%
  \BibitemOpen
  \bibfield  {author} {\bibinfo {author} {\bibfnamefont {A.~Y.}\ \bibnamefont
  {Kitaev}}, \bibinfo {author} {\bibfnamefont {A.~H.}\ \bibnamefont {Shen}},\
  and\ \bibinfo {author} {\bibfnamefont {M.~N.}\ \bibnamefont {Vyalyi}},\
  }\href@noop {} {\emph {\bibinfo {title} {Classical and Quantum
  Computation}}}\ (\bibinfo  {publisher} {American Mathematical Society},\
  \bibinfo {address} {USA},\ \bibinfo {year} {2002})\BibitemShut {NoStop}%
\bibitem [{\citenamefont {Kraus}\ \emph {et~al.}(1983)\citenamefont {Kraus},
  \citenamefont {B{\"o}hm}, \citenamefont {Dollard},\ and\ \citenamefont
  {Wootters}}]{Book:Krauss}%
  \BibitemOpen
  \bibfield  {author} {\bibinfo {author} {\bibfnamefont {K.}~\bibnamefont
  {Kraus}}, \bibinfo {author} {\bibfnamefont {A.}~\bibnamefont {B{\"o}hm}},
  \bibinfo {author} {\bibfnamefont {J.}~\bibnamefont {Dollard}},\ and\ \bibinfo
  {author} {\bibfnamefont {W.}~\bibnamefont {Wootters}},\ }\href
  {https://books.google.pl/books?id=fRBBAQAAIAAJ} {\emph {\bibinfo {title}
  {States, effects, and operations: fundamental notions of quantum theory :
  lectures in mathematical physics at the University of Texas at Austin}}},\
  Lecture notes in physics\ (\bibinfo  {publisher} {Springer-Verlag},\ \bibinfo
  {year} {1983})\BibitemShut {NoStop}%
\bibitem [{\citenamefont {Ruskai]}\ \emph {et~al.}(2002)\citenamefont
  {Ruskai]}, \citenamefont {Szarek},\ and\ \citenamefont
  {Werner}}]{BETHRUSKAI2002159}%
  \BibitemOpen
  \bibfield  {author} {\bibinfo {author} {\bibfnamefont {M.~B.}\ \bibnamefont
  {Ruskai]}}, \bibinfo {author} {\bibfnamefont {S.}~\bibnamefont {Szarek}},\
  and\ \bibinfo {author} {\bibfnamefont {E.}~\bibnamefont {Werner}},\ }\href
  {https://doi.org/https://doi.org/10.1016/S0024-3795(01)00547-X} {\bibfield
  {journal} {\bibinfo  {journal} {Linear Algebra and its Applications}\
  }\textbf {\bibinfo {volume} {347}},\ \bibinfo {pages} {159 } (\bibinfo {year}
  {2002})}\BibitemShut {NoStop}%
\bibitem [{\citenamefont {Streltsov}\ \emph {et~al.}(2018)\citenamefont
  {Streltsov}, \citenamefont {Kampermann}, \citenamefont {Wölk}, \citenamefont
  {Gessner},\ and\ \citenamefont {Bru{\ss}}}]{Streltsov:18}%
  \BibitemOpen
  \bibfield  {author} {\bibinfo {author} {\bibfnamefont {A.}~\bibnamefont
  {Streltsov}}, \bibinfo {author} {\bibfnamefont {H.}~\bibnamefont
  {Kampermann}}, \bibinfo {author} {\bibfnamefont {S.}~\bibnamefont {Wölk}},
  \bibinfo {author} {\bibfnamefont {M.}~\bibnamefont {Gessner}},\ and\ \bibinfo
  {author} {\bibfnamefont {D.}~\bibnamefont {Bru{\ss}}},\ }\href
  {https://doi.org/10.1088/1367-2630/aac484} {\bibfield  {journal} {\bibinfo
  {journal} {New Journal of Physics}\ }\textbf {\bibinfo {volume} {20}},\
  \bibinfo {pages} {053058} (\bibinfo {year} {2018})}\BibitemShut {NoStop}%
\bibitem [{\citenamefont {Obando}\ \emph {et~al.}(2019)\citenamefont {Obando},
  \citenamefont {Passos}, \citenamefont {Paula},\ and\ \citenamefont
  {Huguenin}}]{Passos_Markov}%
  \BibitemOpen
  \bibfield  {author} {\bibinfo {author} {\bibfnamefont {P.~C.}\ \bibnamefont
  {Obando}}, \bibinfo {author} {\bibfnamefont {M.~H.~M.}\ \bibnamefont
  {Passos}}, \bibinfo {author} {\bibfnamefont {F.~M.}\ \bibnamefont {Paula}},\
  and\ \bibinfo {author} {\bibfnamefont {J.~A.~O.}\ \bibnamefont {Huguenin}},\
  }\href {https://doi.org/10.1007/s11128-019-2499-8} {\bibfield  {journal}
  {\bibinfo  {journal} {Quantum Information Processing}\ }\textbf {\bibinfo
  {volume} {19}},\ \bibinfo {pages} {7} (\bibinfo {year} {2019})}\BibitemShut
  {NoStop}%
\bibitem [{\citenamefont {Baumgratz}\ \emph {et~al.}(2014)\citenamefont
  {Baumgratz}, \citenamefont {Cramer},\ and\ \citenamefont
  {Plenio}}]{Baumgratz:14}%
  \BibitemOpen
  \bibfield  {author} {\bibinfo {author} {\bibfnamefont {T.}~\bibnamefont
  {Baumgratz}}, \bibinfo {author} {\bibfnamefont {M.}~\bibnamefont {Cramer}},\
  and\ \bibinfo {author} {\bibfnamefont {M.~B.}\ \bibnamefont {Plenio}},\
  }\href {https://doi.org/10.1103/PhysRevLett.113.140401} {\bibfield  {journal}
  {\bibinfo  {journal} {Phys. Rev. Lett.}\ }\textbf {\bibinfo {volume} {113}},\
  \bibinfo {pages} {140401} (\bibinfo {year} {2014})}\BibitemShut {NoStop}%
\bibitem [{\citenamefont {Streltsov}\ \emph {et~al.}(2017)\citenamefont
  {Streltsov}, \citenamefont {Adesso},\ and\ \citenamefont
  {Plenio}}]{Streltsov:17}%
  \BibitemOpen
  \bibfield  {author} {\bibinfo {author} {\bibfnamefont {A.}~\bibnamefont
  {Streltsov}}, \bibinfo {author} {\bibfnamefont {G.}~\bibnamefont {Adesso}},\
  and\ \bibinfo {author} {\bibfnamefont {M.~B.}\ \bibnamefont {Plenio}},\
  }\href {https://doi.org/10.1103/RevModPhys.89.041003} {\bibfield  {journal}
  {\bibinfo  {journal} {Rev. Mod. Phys.}\ }\textbf {\bibinfo {volume} {89}},\
  \bibinfo {pages} {041003} (\bibinfo {year} {2017})}\BibitemShut {NoStop}%
\bibitem [{\citenamefont {Jones}(1941)}]{Jones-i:41}%
  \BibitemOpen
  \bibfield  {author} {\bibinfo {author} {\bibfnamefont {R.~C.}\ \bibnamefont
  {Jones}},\ }\href {https://doi.org/10.1364/JOSA.31.000488} {\bibfield
  {journal} {\bibinfo  {journal} {J. Opt. Soc. Am.}\ }\textbf {\bibinfo
  {volume} {31}},\ \bibinfo {pages} {488} (\bibinfo {year} {1941})}\BibitemShut
  {NoStop}%
\bibitem [{\citenamefont {Simon}\ and\ \citenamefont
  {Mukunda}(1990)}]{Simon:90}%
  \BibitemOpen
  \bibfield  {author} {\bibinfo {author} {\bibfnamefont {R.}~\bibnamefont
  {Simon}}\ and\ \bibinfo {author} {\bibfnamefont {N.}~\bibnamefont
  {Mukunda}},\ }\href
  {https://doi.org/https://doi.org/10.1016/0375-9601(90)90732-4} {\bibfield
  {journal} {\bibinfo  {journal} {Physics Letters A}\ }\textbf {\bibinfo
  {volume} {143}},\ \bibinfo {pages} {165} (\bibinfo {year}
  {1990})}\BibitemShut {NoStop}%
\bibitem [{\citenamefont {Reddy}\ \emph {et~al.}(2014)\citenamefont {Reddy},
  \citenamefont {Prabhakar}, \citenamefont {Aadhi}, \citenamefont {Kumar},
  \citenamefont {Shah}, \citenamefont {Singh},\ and\ \citenamefont
  {Simon}}]{Reddy:14}%
  \BibitemOpen
  \bibfield  {author} {\bibinfo {author} {\bibfnamefont {S.~G.}\ \bibnamefont
  {Reddy}}, \bibinfo {author} {\bibfnamefont {S.}~\bibnamefont {Prabhakar}},
  \bibinfo {author} {\bibfnamefont {A.}~\bibnamefont {Aadhi}}, \bibinfo
  {author} {\bibfnamefont {A.}~\bibnamefont {Kumar}}, \bibinfo {author}
  {\bibfnamefont {M.}~\bibnamefont {Shah}}, \bibinfo {author} {\bibfnamefont
  {R.~P.}\ \bibnamefont {Singh}},\ and\ \bibinfo {author} {\bibfnamefont
  {R.}~\bibnamefont {Simon}},\ }\href {https://doi.org/10.1364/JOSAA.31.000610}
  {\bibfield  {journal} {\bibinfo  {journal} {J. Opt. Soc. Am. A}\ }\textbf
  {\bibinfo {volume} {31}},\ \bibinfo {pages} {610} (\bibinfo {year}
  {2014})}\BibitemShut {NoStop}%
\bibitem [{\citenamefont {Huang}\ \emph {et~al.}(2018)\citenamefont {Huang},
  \citenamefont {Wang}, \citenamefont {Rohde}, \citenamefont {Luo},
  \citenamefont {Zhao}, \citenamefont {Liu}, \citenamefont {Li}, \citenamefont
  {Liu}, \citenamefont {Lu},\ and\ \citenamefont {Pan}}]{Huang18}%
  \BibitemOpen
  \bibfield  {author} {\bibinfo {author} {\bibfnamefont {H.-L.}\ \bibnamefont
  {Huang}}, \bibinfo {author} {\bibfnamefont {X.-L.}\ \bibnamefont {Wang}},
  \bibinfo {author} {\bibfnamefont {P.~P.}\ \bibnamefont {Rohde}}, \bibinfo
  {author} {\bibfnamefont {Y.-H.}\ \bibnamefont {Luo}}, \bibinfo {author}
  {\bibfnamefont {Y.-W.}\ \bibnamefont {Zhao}}, \bibinfo {author}
  {\bibfnamefont {C.}~\bibnamefont {Liu}}, \bibinfo {author} {\bibfnamefont
  {L.}~\bibnamefont {Li}}, \bibinfo {author} {\bibfnamefont {N.-L.}\
  \bibnamefont {Liu}}, \bibinfo {author} {\bibfnamefont {C.-Y.}\ \bibnamefont
  {Lu}},\ and\ \bibinfo {author} {\bibfnamefont {J.-W.}\ \bibnamefont {Pan}},\
  }\href {https://doi.org/10.1364/OPTICA.5.000193} {\bibfield  {journal}
  {\bibinfo  {journal} {Optica}\ }\textbf {\bibinfo {volume} {5}},\ \bibinfo
  {pages} {193} (\bibinfo {year} {2018})}\BibitemShut {NoStop}%
\bibitem [{\citenamefont {Huang}\ \emph {et~al.}(2017)\citenamefont {Huang},
  \citenamefont {Zhao}, \citenamefont {Ma}, \citenamefont {Liu}, \citenamefont
  {Su}, \citenamefont {Wang}, \citenamefont {Li}, \citenamefont {Liu},
  \citenamefont {Sanders}, \citenamefont {Lu},\ and\ \citenamefont
  {Pan}}]{Huang17}%
  \BibitemOpen
  \bibfield  {author} {\bibinfo {author} {\bibfnamefont {H.-L.}\ \bibnamefont
  {Huang}}, \bibinfo {author} {\bibfnamefont {Q.}~\bibnamefont {Zhao}},
  \bibinfo {author} {\bibfnamefont {X.}~\bibnamefont {Ma}}, \bibinfo {author}
  {\bibfnamefont {C.}~\bibnamefont {Liu}}, \bibinfo {author} {\bibfnamefont
  {Z.-E.}\ \bibnamefont {Su}}, \bibinfo {author} {\bibfnamefont {X.-L.}\
  \bibnamefont {Wang}}, \bibinfo {author} {\bibfnamefont {L.}~\bibnamefont
  {Li}}, \bibinfo {author} {\bibfnamefont {N.-L.}\ \bibnamefont {Liu}},
  \bibinfo {author} {\bibfnamefont {B.~C.}\ \bibnamefont {Sanders}}, \bibinfo
  {author} {\bibfnamefont {C.-Y.}\ \bibnamefont {Lu}},\ and\ \bibinfo {author}
  {\bibfnamefont {J.-W.}\ \bibnamefont {Pan}},\ }\href
  {https://doi.org/10.1103/PhysRevLett.119.050503} {\bibfield  {journal}
  {\bibinfo  {journal} {Phys. Rev. Lett.}\ }\textbf {\bibinfo {volume} {119}},\
  \bibinfo {pages} {050503} (\bibinfo {year} {2017})}\BibitemShut {NoStop}%
\bibitem [{\citenamefont {Lu}\ \emph {et~al.}(2007)\citenamefont {Lu},
  \citenamefont {Browne}, \citenamefont {Yang},\ and\ \citenamefont
  {Pan}}]{Lu07}%
  \BibitemOpen
  \bibfield  {author} {\bibinfo {author} {\bibfnamefont {C.-Y.}\ \bibnamefont
  {Lu}}, \bibinfo {author} {\bibfnamefont {D.~E.}\ \bibnamefont {Browne}},
  \bibinfo {author} {\bibfnamefont {T.}~\bibnamefont {Yang}},\ and\ \bibinfo
  {author} {\bibfnamefont {J.-W.}\ \bibnamefont {Pan}},\ }\href
  {https://doi.org/10.1103/PhysRevLett.99.250504} {\bibfield  {journal}
  {\bibinfo  {journal} {Phys. Rev. Lett.}\ }\textbf {\bibinfo {volume} {99}},\
  \bibinfo {pages} {250504} (\bibinfo {year} {2007})}\BibitemShut {NoStop}%
\bibitem [{\citenamefont {Sasada}\ and\ \citenamefont
  {Okamoto}(2003)}]{MZIM_Sassada}%
  \BibitemOpen
  \bibfield  {author} {\bibinfo {author} {\bibfnamefont {H.}~\bibnamefont
  {Sasada}}\ and\ \bibinfo {author} {\bibfnamefont {M.}~\bibnamefont
  {Okamoto}},\ }\href {https://doi.org/10.1103/PhysRevA.68.012323} {\bibfield
  {journal} {\bibinfo  {journal} {Phys. Rev. A}\ }\textbf {\bibinfo {volume}
  {68}},\ \bibinfo {pages} {012323} (\bibinfo {year} {2003})}\BibitemShut
  {NoStop}%
\bibitem [{\citenamefont {Altepeter}\ \emph {et~al.}(2005)\citenamefont
  {Altepeter}, \citenamefont {Jeffrey},\ and\ \citenamefont
  {Kwiat}}]{Photonic-State-Tomography}%
  \BibitemOpen
  \bibfield  {author} {\bibinfo {author} {\bibfnamefont {J.}~\bibnamefont
  {Altepeter}}, \bibinfo {author} {\bibfnamefont {E.}~\bibnamefont {Jeffrey}},\
  and\ \bibinfo {author} {\bibfnamefont {P.}~\bibnamefont {Kwiat}}\ }(\bibinfo
  {publisher} {Academic Press},\ \bibinfo {year} {2005})\ pp.\ \bibinfo {pages}
  {105 -- 159}\BibitemShut {NoStop}%
\bibitem [{\citenamefont {Passos}\ \emph {et~al.}(2019)\citenamefont {Passos},
  \citenamefont {Santos}, \citenamefont {Sarandy},\ and\ \citenamefont
  {Huguenin}}]{Passos_Machine}%
  \BibitemOpen
  \bibfield  {author} {\bibinfo {author} {\bibfnamefont {M.~H.~M.}\
  \bibnamefont {Passos}}, \bibinfo {author} {\bibfnamefont {A.~C.}\
  \bibnamefont {Santos}}, \bibinfo {author} {\bibfnamefont {M.~S.}\
  \bibnamefont {Sarandy}},\ and\ \bibinfo {author} {\bibfnamefont {J.~A.~O.}\
  \bibnamefont {Huguenin}},\ }\href@noop {} {\bibfield  {journal} {\bibinfo
  {journal} {Phys. Rev. A}\ }\textbf {\bibinfo {volume} {100}},\ \bibinfo
  {pages} {022113} (\bibinfo {year} {2019})}\BibitemShut {NoStop}%
\bibitem [{\citenamefont {Gopaul}\ and\ \citenamefont
  {Andrews}(2007)}]{Gopaul:2007}%
  \BibitemOpen
  \bibfield  {author} {\bibinfo {author} {\bibfnamefont {C.}~\bibnamefont
  {Gopaul}}\ and\ \bibinfo {author} {\bibfnamefont {R.}~\bibnamefont
  {Andrews}},\ }\href {https://doi.org/10.1088/1367-2630/9/4/094} {\bibfield
  {journal} {\bibinfo  {journal} {New Journal of Physics}\ }\textbf {\bibinfo
  {volume} {9}},\ \bibinfo {pages} {94} (\bibinfo {year} {2007})}\BibitemShut
  {NoStop}%
\bibitem [{\citenamefont {Ge}\ \emph {et~al.}(2015)\citenamefont {Ge},
  \citenamefont {Wang},\ and\ \citenamefont {Guo}}]{Ge:2015}%
  \BibitemOpen
  \bibfield  {author} {\bibinfo {author} {\bibfnamefont {X.-L.}\ \bibnamefont
  {Ge}}, \bibinfo {author} {\bibfnamefont {B.-Y.}\ \bibnamefont {Wang}},\ and\
  \bibinfo {author} {\bibfnamefont {C.-S.}\ \bibnamefont {Guo}},\ }\href
  {https://doi.org/10.1364/JOSAA.32.000837} {\bibfield  {journal} {\bibinfo
  {journal} {J. Opt. Soc. Am. A}\ }\textbf {\bibinfo {volume} {32}},\ \bibinfo
  {pages} {837} (\bibinfo {year} {2015})}\BibitemShut {NoStop}%
\bibitem [{\citenamefont {Lochab}\ \emph {et~al.}(2017)\citenamefont {Lochab},
  \citenamefont {Senthilkumaran},\ and\ \citenamefont {Khare}}]{Lochab:2017}%
  \BibitemOpen
  \bibfield  {author} {\bibinfo {author} {\bibfnamefont {P.}~\bibnamefont
  {Lochab}}, \bibinfo {author} {\bibfnamefont {P.}~\bibnamefont
  {Senthilkumaran}},\ and\ \bibinfo {author} {\bibfnamefont {K.}~\bibnamefont
  {Khare}},\ }\href {https://doi.org/10.1364/OE.25.017524} {\bibfield
  {journal} {\bibinfo  {journal} {Opt. Express}\ }\textbf {\bibinfo {volume}
  {25}},\ \bibinfo {pages} {17524} (\bibinfo {year} {2017})}\BibitemShut
  {NoStop}%
\end{thebibliography}%
\end{document}